\documentclass[preprintnumbers,twocolumn,amsmath,amssymb,aps,nofootinbib,groupedaddress]{revtex4-2}

\usepackage{graphicx}
\usepackage{dcolumn}
\usepackage{bm}
\usepackage{hyperref}
\hypersetup{colorlinks,allcolors=black}

\usepackage{tikz}

\usepackage{wrapfig, blindtext}
\usepackage{arydshln}
\usepackage{pgfplots}
\usepackage{axodraw2}
\pgfplotsset{compat=1.17}
\usepackage{systeme}
\usepackage{tikz}
\usetikzlibrary{positioning,arrows,arrows.meta,shapes}
\usepackage{tcolorbox}
\usepackage{xcolor}
\usepackage{listings}
\usepackage{tikz-3dplot}
\usetikzlibrary{decorations.markings}
\usetikzlibrary{intersections}
\usepackage{capt-of}
\usepackage{comment}

\usetikzlibrary{spath3, intersections, hobby}

\usepackage{scalerel,stackengine}
\stackMath

\newcommand\reallywidehat[1]{%
  \savestack{\tmpbox}{\stretchto{%
    \scaleto{%
      \scalerel*[\widthof{\ensuremath{#1}}]{\kern-.6pt\bigwedge\kern-.6pt}%
      {\rule[-\textheight/2]{1ex}{\textheight}}
    }{\textheight}%
  }{0.5ex}}%
  \stackon[1pt]{#1}{\tmpbox}%
}

\usepackage{tikz-cd}

\newcommand{\Braket}[2]{\left \langle #1 \middle \rvert #2 \right \rangle}

\newcommand{\kket}[1]{\ensuremath{|#1\rangle\rangle}}
\newcommand{\bbra}[1]{\ensuremath{\langle\langle#1|}}
\newcommand{\ketbra}[2]{\mathinner{\left|#1\right>\!\left<#2\right|}}

\usepackage{graphicx}
\usepackage{slashed}
\usepackage{tabularx,ragged2e}
\usepackage{amssymb,fge}
\usepackage{amsmath,amssymb}
\usepackage{slashed}
\usepackage{hyperref}
\usepackage{caption}
\usepackage{dsfont}
\usepackage{verbatim}
\usepackage{subfig}
\usepackage{mathtools,ytableau,amsfonts,tikz}
\usepackage{graphicx}
\usepackage{physics}
\usepackage{amsthm}
\usepackage{tcolorbox}
\usepackage[normalem]{ulem}
\usepackage{mathrsfs}

\begin{document}


\title{Unimodular JT gravity and de Sitter quantum cosmology}
\author{Bruno Alexandre$^1$}
\email{bruno.alexandre20@imperial.ac.uk}
\author{Altay Etkin$^2$}
\email{a.etkin@soton.ac.uk}
\author{Farbod-Sayyed Rassouli$^3$}
\email{sayyed.rassouli@nottingham.ac.uk}

\affiliation{$^{1}$Abdus Salam Centre for Theoretical Physics, Imperial College London, Prince Consort Road, London, SW7 2AZ, UK}

\affiliation{$^{2}$School of Mathematical Sciences and STAG Research Centre, University of Southampton, Highfield Campus, Southampton, SO17 1BJ, UK}

\affiliation{$^{3}$School of Physics and Astronomy and Nottingham Centre of Gravity, University of Nottingham, University Park, Nottingham, NG7 2RD, UK}

\date{12 March 2025}

\begin{abstract}
In this work, we show that a gauge-theoretic description of Jackiw-Teitelboim (JT) gravity naturally yields a Henneaux-Teitelboim (HT) unimodular gravity via a central extension of its isometry group, valid for both flat and curved two-dimensional spacetimes. HT gravity introduces a unimodular time canonically conjugate to the cosmological constant, serving as a physical time in quantum cosmology. By studying the mini-superspace reduction of HT$_2$ gravity, the Wheeler-DeWitt equation becomes a Schr\"odinger-like equation, giving a consistent and unitary quantum theory. Analysis of the wavefunction's probability density reveals a quantum distribution for the scale factor \(a\), offering a quantum perspective on the expansion and contraction of the universe. In this perspective, the possibility of reaching the singular point \(a=0\) signals that topology change could occur. Finally, we give a consistent quantum description of unimodular time that aligns seamlessly with Page-Wootters formulation of quantum mechanics, where quantum correlations between unimodular time and JT gravity are studied in HT$_2$ quantum cosmology.
\end{abstract}

\maketitle

\section{Introduction}
One of the profound challenges in theoretical physics is developing a quantum-mechanical framework that encompasses the entire universe. Indeed, the incompatibility between General Relativity (GR) and Quantum Mechanics (QM) is particularly evident in the well-known problem of time, which arises due to the fundamentally different treatment of time in both theories \cite{Kuchar:1991xd,Bombelli:1991jj,Daughton:1993uy,sorkin1,Daughton:1998aa}. In GR, time is treated as part of the spacetime fabric and therefore, the theory is diffeomorphism invariant. Conversely, in QM time is treated as an external, absolute parameter used to measure the evolution of a quantum system. Indeed, the Hamiltonian constraint of a gravitational system vanishes, therefore leading to the absence of unitary evolution in terms of quantum states.

A more recent and promising avenue is Henneaux-Teitelboim (HT) gravity where by converting the constants of nature to dynamical variables one gets physical relational times which are the canonical conjugates of the dynamical “constants” \cite{Einstein-unimod,Weinberg:1988cp,Henneaux:1989zc,unimod1,Smolin:2009ti,Smolin:2010iq,Kaloper:2013zca,FernandezCristobal:2014jca,Padilla:2014yea,Bufalo:2015wda,pad1,Percacci:2017fsy,lombriser2019cosmological,Carballo-Rubio:2022ofy}. Therefore we end up having a physical time variable conjugate to a dynamical cosmological constant \cite{Alexander:2018djy,Magueijo:2020ntm,Magueijo:2021rpi,Magueijo:2021pvq,Jirousek:2018ago,Jirousek:2020vhy,Vikman:2021god}, which preserves unitarity and allows the construction of normalizable wave packets \cite{Alexandre:2022npo,Alexandre:2023ozf,PathInt}. This not only converts the Wheeler-DeWitt equation into a Schr\"{o}dinger-like equation, but also resolves the apparent unitarity problem and gives conserved probabilities.

In this regard, we study JT gravity as a two-dimensional toy model for quantum gravity and quantum cosmology. Ordinarily, in JT gravity the Hamiltonian constraint is found to vanish, resulting in “frozen” quantum states. In turn, this exacerbates the problem of the absence of unitary time evolution. The missing of an external time in (de Sitter) JT gravity has been extensively discussed in the context of the CLPW framework of algebras of observables \cite{Chandrasekaran:2022cip, Witten:2023qsv, Gomez:2023wrq, Gomez:2023upk, Witten:2023xze}. In this framework, observers are treated as clocks within the gravitating system, whose role is to fix the time translation symmetry. Put simply, an observer is any system capable of measuring time. Recently, it has also been demonstrated that the CLPW algebras of observables is equivalent to the Page-Wootters (PW) approach \cite{Page:1983uc,Wootters:1984wfv,Giovannetti:2015qha,Smith:2017pwx,Hoehn:2019fsy,DeVuyst:2024pop,DeVuyst:2024uvd} where the total system is described by correlations between the observer and the gravitational system.

In this paper, we show that a physical clock naturally emerges from a HT formulation of JT gravity, which in turn stems from a gauge-theoretic approach—a connection previously unrecognized in the literature (such as the $\reallywidehat{\text{\textcolor{black}{CGHS}}}$ model discussed in two-dimensional flat holography) \cite{Cangemi:1992bj,Jackiw:1992ev,Cangemi:1992ri,Grignani:1992hw,Cangemi:1992up,Kim:1992fb,Kim:1992ht,Cangemi:1993sd,Cangemi:1993bb,Jackiw:1993gf}. In this scenario, introducing unimodular time allows the PW formalism to reach its full potential, providing a robust framework in which a genuine notion of time evolution can be defined. Moreover, the ability to define such a physical clock in a controlled, lower-dimensional model of gravity opens new avenues for studying quantum cosmology. In particular, it is especially enticing that for de Sitter space, one can study time evolution of closed universes and topology change at early times.

The paper is organized as follows: in section \ref{sec:HTcentralextension}, we show how HT gravity naturally emerges from a gauge formulation of JT gravity via central extensions of the isometry group. In section \ref{sec:qcHT2}, we explore this formulation in mini-superspace (MSS), where we quantize the theory and build normalizable unimodular wave packets. We study the probability density associated with such wave packets in section \ref{sec:topologychange} and comment on the global topology of the spacetime. In section \ref{sec:PW}, we introduce the Page-Wootters formalism of QM and apply it to the work done in the previous sections. We conclude with some final remarks in section \ref{sec:conclusion}.

\section{Henneaux-Teitelboim gravity from central extensions of JT gravity}\label{sec:HTcentralextension}

It is well established that GR admits a volume-preserving diffeomorphism-invariant formulation, commonly referred to as unimodular gravity \cite{Einstein-unimod,Weinberg:1988cp,Henneaux:1989zc,unimod1,Smolin:2009ti,Smolin:2010iq,Kaloper:2013zca,FernandezCristobal:2014jca,Padilla:2014yea,Bufalo:2015wda,pad1,Percacci:2017fsy,lombriser2019cosmological,Carballo-Rubio:2022ofy}. This formulation is classically equivalent to GR and is achieved by imposing the gauge fixing condition $\sqrt{-g} = 1$. The seminal work of Henneaux and Teitelboim extended this framework by restoring full diffeomorphism invariance in unimodular gravity through the promotion of constants of nature to dynamical variables \cite{Alexander:2018djy,Magueijo:2020ntm,Magueijo:2021rpi,Magueijo:2021pvq,Jirousek:2018ago,Jirousek:2020vhy,Vikman:2021god}. This parametrization becomes evident when the unimodular condition is modified to $\sqrt{-g} = \partial_\mu \mathcal{T}^\mu$, treated as an on-shell expression, with action given by
\begin{equation}
    S = \frac{1}{2} \int d^4x \ \sqrt{-g}(R-2 \Lambda)+\int d^4x \ \Lambda \partial_\mu \mathcal{T}^\mu,
\end{equation}
where $\mathcal{T}^\mu$ is an auxiliary vector density, and $\Lambda$ is the vacuum energy field (also known as the cosmological constant) promoted to an off-shell variable, and on-shell constant. This theory yields equations of motion identical to those of the standard Einstein–Hilbert action. 

Recent work \cite{Hammer:2020dqp,Etkin:2023amf} has demonstrated that an abelian reduction based on the gauge group $\text{U}(1)$ yields an HT gravity theory, where the HT term is replaced by the Chern-Pontryagin topological term. The resultant action is
\begin{equation}\label{eq:U(1)HT4action}
    S = \frac{1}{2} \int d^4x \ \sqrt{-g} \bigl(R-2 \Lambda\bigr) - \frac{1}{8 \pi^2} \int \Lambda F \wedge F,
\end{equation}
with $F=dA$ the curvature two-form associated with an abelian $\text{U}(1)$ gauge field $A$, and with the vector density $\mathcal{T}^{\mu}$ identified as the Chern-Simons current.

Although typically studied in four dimensions, HT gravity also emerges in certain two-dimensional contexts, though not recognized as such. In particular, various formulations of JT gravity—such as those employed in flat holography \cite{Gonzalez:2018enk,Afshar:2019tvp,Afshar:2019axx,Godet:2021cdl,Afshar:2021qvi,Kar:2022sdc,Rosso:2022tsv,Afshar:2022mkf}—are revealed to be HT gravity theories \cite{Callan:1992rs,Cangemi:1992bj,Jackiw:1992ev,Cangemi:1992ri,Grignani:1992hw,Cangemi:1992up,Kim:1992fb,Kim:1992ht,Cangemi:1993sd,Cangemi:1993bb,Jackiw:1993gf,Godet:2020xpk} in this paper.

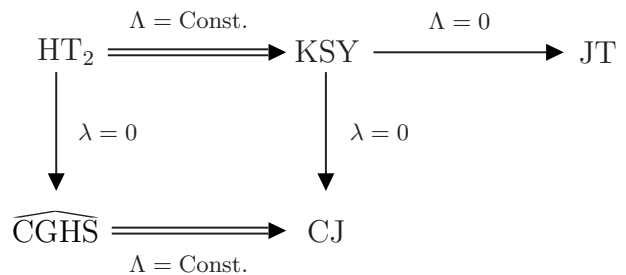
\begin{figure}[htbp]
    \centering
    \tikzset{every picture/.style={line width=0.75pt}} 
    
    \begin{tikzpicture}[x=0.75pt,y=0.75pt,yscale=-1,xscale=1]
    
    \draw    (60,48.5) -- (141,48.5)(60,51.5) -- (141,51.5) ;
    \draw [shift={(150,50)}, rotate = 180] [fill={rgb, 255:red, 0; green, 0; blue, 0 }  ][line width=0.08]  [draw opacity=0] (8.93,-4.29) -- (0,0) -- (8.93,4.29) -- cycle    ;
    \draw    (62,138.5) -- (100.87,138.5) -- (141,138.5)(62,141.5) -- (100.87,141.5) -- (141,141.5) ;
    \draw [shift={(150,140)}, rotate = 180] [fill={rgb, 255:red, 0; green, 0; blue, 0 }  ][line width=0.08]  [draw opacity=0] (8.93,-4.29) -- (0,0) -- (8.93,4.29) -- cycle    ;
    \draw    (34,60) -- (34,117) ;
    \draw [shift={(34,120)}, rotate = 270] [fill={rgb, 255:red, 0; green, 0; blue, 0 }  ][line width=0.08]  [draw opacity=0] (8.93,-4.29) -- (0,0) -- (8.93,4.29) -- cycle    ;
    \draw    (170,60) -- (170,117) ;
    \draw [shift={(170,120)}, rotate = 270] [fill={rgb, 255:red, 0; green, 0; blue, 0 }  ][line width=0.08]  [draw opacity=0] (8.93,-4.29) -- (0,0) -- (8.93,4.29) -- cycle    ;
    \draw    (194,50) -- (287,50) ;
    \draw [shift={(290,50)}, rotate = 180] [fill={rgb, 255:red, 0; green, 0; blue, 0 }  ][line width=0.08]  [draw opacity=0] (8.93,-4.29) -- (0,0) -- (8.93,4.29) -- cycle    ;
    
    \draw (39,49.5) node  [font=\large]  {$\text{HT}_{2}$};
    \draw (33.67,137.87) node  [font=\large]  {$\reallywidehat{\text{\textcolor{black}{CGHS}}}$};
    \draw (60,90) node    {$\lambda =0$};
    \draw (170.5,49.5) node  [font=\large] [align=left] {KSY};
    \draw (197,90) node    {$\lambda =0$};
    \draw (100.85,33.87) node    {$\Lambda =\text{Const.}$};
    \draw (100.85,156.87) node    {$\Lambda =\text{Const.}$};
    \draw (237.42,34) node    {$\Lambda =0$};
    \draw (296,43) node [anchor=north west][inner sep=0.75pt]  [font=\large] [align=left] {JT};
    \draw (170.17,140.43) node  [font=\large] [align=left] {CJ};

    \end{tikzpicture}
    \caption{\textcolor{black}{Diagram illustrating various reductions of the general $\text{HT}_2$ gravity theory. The double arrows indicate the gauge formulations of the KSY and CJ gravity models.}}
    \label{fig:gaugeformulations}
\end{figure}

This structure arises naturally when JT gravity is formulated from a gauge-theoretic perspective (see figure \ref{fig:gaugeformulations}). The standard JT action, without the inclusion of boundary terms, is given by
\begin{equation}\label{eq:JT action}
S_{\text{JT}} = \frac{1}{2} \int d^2x \ \sqrt{-g}\phi\bigl(R -2\lambda\bigr),
\end{equation}
where $\lambda$ is the ``cosmological constant'' that ensures that upon variation of the dilaton field $\phi$, we have the on-shell expression $R=2\lambda$. The dilaton equation of motion is obtained from the metric variation of the action:
\begin{equation}
    \bigl(\nabla_{\mu} \partial_{\nu} - g_{\mu \nu} \Box\bigr)\phi = \lambda g_{\mu \nu} \phi.
\end{equation}
The global spacetime geometry depends on the value of $\lambda$: it is de Sitter (dS) for $\lambda > 0$ and anti-de Sitter (AdS) for $\lambda < 0$. 

Of particular interest is the JT gravity action \eqref{eq:JT action} with non-vanishing $\lambda$, which possesses a gauge theoretical formulation based on the $\mathfrak{sl}(2,\mathbb{R})$-algebra \cite{Teitelboim:1983ux,Jackiw:1984je,Isler:1989hq,Chamseddine:1989yz}:
\begin{align}
    \left[P_a, P_b\right] &=2 \lambda J_{a b} \equiv-\lambda \epsilon_{a b} \widetilde{J},\nonumber\\
 [P_a, \widetilde{J}] &=-\epsilon_a^{\ b} P_b.
\end{align}
The resulting first-order formulation is a BF theory with action given by
\begin{equation}
    S = \int \bigl\langle \mathbf{B},\mathbf{F} \bigr\rangle = \int B_a T^a + \widetilde{B} \bigl(\widetilde{R}-\lambda \widetilde{\Sigma}\bigr),
\end{equation}
where $\mathbf{B} = B^a P_a + \widetilde{B} \widetilde{J}$ is a zero-form Lagrange multiplier, and 
\begin{equation}
    \mathbf{F} = d\mathbf{A} + \mathbf{A} \wedge \mathbf{A} = T^a P_a + \widetilde{R} \widetilde{J},
\end{equation}
is the curvature two-form associated with the connection one-form $\mathbf{A} = e^a P_a + \widetilde{\omega} \widetilde{J}$ for the $\mathfrak{sl}(2,\mathbb{R})$-algebra. Here, the Einstein-Cartan (EC) variables are the zweibein $e^a$ and dual spin-connection $\widetilde{\omega} = \frac{1}{2} \epsilon^{ab} \omega_{ab}$, with associated torsion $T^a = De^a$ and dual curvature $\widetilde{R} = \frac{1}{2} \epsilon^{ab} R_{ab} = d\widetilde{\omega}$. The dual area two-form is $\widetilde{\Sigma} = \frac{1}{2} \epsilon_{ab} \Sigma^{ab}$, where $\Sigma^{ab} = e^a \wedge e^b$. In this context, one can show that the on-shell action reduces to an Einstein-Cartan action for JT gravity 
\begin{equation}
    S_{\text{JT}} = \int \phi \bigl(\widetilde{R}-\lambda \widetilde{\Sigma}\bigr),
\end{equation}
with vanishing torsion imposed by the Lagrange multiplier $B^a$ and $\widetilde{B}$ identified with the dilaton $\phi$.

Building on top of the requirements for an HT gravity theory, one can formulate a two-dimensional unimodular JT gravity model ($\text{HT}_2$), described by the action:
\begin{equation}\label{eq:HT2 action}
S_{\text{HT}_2} :=\frac{1}{2} \int d^2x \ \sqrt{-g}\Bigl(\phi\bigl(R -2\lambda\bigr)-2\Lambda\Bigr) + \int d^2x \ \Lambda \partial_\mu \mathcal{T}^\mu.
\end{equation}
The distinction between $\lambda$ and $\Lambda$ is crucial: $\lambda$ determines the global spacetime geometry on-shell, while $\Lambda$ serves as the vacuum energy field, similar to its role in four-dimensional HT gravity. In this section, we demonstrate that the action \eqref{eq:HT2 action} emerges naturally from a gauge-invariant formulation of JT gravity and study its features.

\subsection{$\text{HT}_2$ as a gauge-invariant formulation of JT gravity}

We want to show that a gauge formulation of \eqref{eq:HT2 action} arises from a BF formulation, akin to JT gravity. For that, we follow Kim, Soh, and Yee (KSY) \cite{Kim:1992fb,Kim:1992ht}, where they consider an extension of JT gravity via a vacuum cosmological constant term:
\begin{equation}\label{eq:eqksy}
    S_{\text{KSY}} = \frac{1}{2} \int d^2x \ \sqrt{-g} \Bigl(\phi \bigl(R - 2\lambda\bigr) - 2 \Lambda\Bigr).
\end{equation}
The on-shell expressions are obtain by varying the action with respect to $\phi$ and $g_{\mu \nu}$, respectively:
\begin{align}
    R &= 2\lambda,\\
\bigl(\nabla_{\mu} \partial_{\nu} - g_{\mu \nu} \Box\bigr)\phi &= (\lambda\phi+\Lambda) g_{\mu \nu}\label{eq:varg}.
\end{align}

A gauge invariant formulation of this model with $\lambda \neq 0$ is based on an abelian central extension of the isometry group of spacetime, that is a BF theory with Lie algebra: 
\begin{equation}
    \reallywidehat{\mathfrak{sl}}(2,\mathbb{R}) \cong \mathfrak{sl}(2,\mathbb{R})\oplus \mathfrak{u}(1),
\end{equation}
and commutation relations given by:
\begin{align}\label{eq:commutationsl(2,R)}
 \left[P_a, P_b\right] &= -\lambda \epsilon_{a b} \widetilde{J} + \epsilon_{ab} I,\\
 [P_a, \widetilde{J}] &=-\epsilon_a^{\ b} P_b,
\end{align}
where $I$ is the central element given by the $\mathfrak{u}(1)$ subalgebra. 
With this algebra, one can construct an $\reallywidehat{\mathfrak{sl}}(2,\mathbb{R})$-valued connection one-form:
\begin{align}
    \mathbf{A}=e^aP_a +\widetilde{\omega}\widetilde{J}+AI, 
\end{align}
and associated curvature two-form given by
\begin{align}
    \mathbf{F}=T^a P_a +(\widetilde{R}-\lambda \widetilde{\Sigma}) \widetilde{J}+(F+\widetilde{\Sigma})I.
\end{align}
Here, $F = dA$ is the $\mathfrak{u}(1)$-valued curvature two-form with  associated connection one-form $A$. Using the above curvature $\mathbf{F}$ and defining the Lagrange multiplier zero-form $\mathbf{B} = B^a P_a + \widetilde{B} \widetilde{J} + B I$, the BF action reduces to
\begin{equation}\label{eq:BF}
    S = \int B_a T^a +\phi( \widetilde{R}- \lambda \widetilde{\Sigma}) - \Lambda(F+\widetilde{\Sigma}),
\end{equation}
where we set $B=\phi+\Lambda$ and $\widetilde{B} = -\lambda \phi - \Lambda$ using the trace identities induced from the Casimir invariant (see \cite{Cangemi:1993bb}). The vanishing of torsion gives the on-shell action, corresponding to an EC formulation of a gauge invariant first-order action for the KSY model given by:
\begin{equation}\label{eq:EC HT2 action}
    S_{\text{HT}_2} = \int \phi( \widetilde{R}- \lambda \widetilde{\Sigma}) - \Lambda(F+\widetilde{\Sigma}),
\end{equation}
Transitioning to the second-order formulation, we recover the same action as in \eqref{eq:HT2 action}, which is:
\begin{equation}\label{eq:HT2actionA}
    S_{\text{HT}_2} = \frac{1}{2}\int d^2x \ \sqrt{-g}\phi(R -2\lambda) + \int d^2x \ \sqrt{-g} \Lambda\bigl(\epsilon^{\mu \nu} \partial_{\mu} A_{\nu} - 1\bigr),
\end{equation}
where the vector density $\mathcal{T}^\mu$ is identified with the dual of the abelian gauge field, $\mathcal{A}^\mu = \varepsilon^{\mu \nu} A_\nu$. Here, $\varepsilon^{\mu \nu}$ denotes the Levi-Civita symbol.\footnote{We denote $\epsilon^{\mu \nu}$ by the Levi-Civita tensor given by $\epsilon^{\mu \nu} = \frac{\varepsilon^{\mu \nu}}{\sqrt{-g}}$.}

The variation of the action with respect to $\phi$, $g_{\mu\nu}$, $\mathcal{A}^\mu$ and $\Lambda$, respectively, yields the following equations of motion: 
\begin{align}
    &R = 2\lambda, \label{eq:varphi}\\[4pt]
    &\bigl(\nabla_{\mu} \partial_{\nu} - g_{\mu \nu} \Box\bigr)\phi = (\lambda\phi+\Lambda) g_{\mu \nu},\label{eq:varg} \\[4pt]
    &\partial_\mu \Lambda = 0,\label{eq:varA}\\[4pt]
    &\sqrt{-g} = \partial_\mu \mathcal{A}^\mu. \label{eq:varLambda}
\end{align}
From the first three equations of motion, the on-shell dynamics of \(\text{HT}_2\) match those of the KSY model, provided that the unimodular condition \eqref{eq:varLambda} is satisfied. Therefore, the classical dynamics remain unchanged and they describe the same physics \cite{Henneaux:1989zc, Padilla:2014yea}. Indeed, $\Lambda$ is promoted to a phase-space variable, canonically conjugate to the zero-mode of $\mathcal{A}^0$, which is referred to as unimodular time \cite{Henneaux:1989zc}
\begin{equation}
T_{\Lambda}(t):=\int_{\Sigma_t} d\Sigma \  \mathcal{A}^0 \Big|_{\Sigma_t} = -\int_{\Sigma_t} d\Sigma \ A_1 \Big|_{\Sigma_t},
\end{equation}
where $\Sigma_t$ are constant time hypersurfaces and $\mathcal{A}^0 = -A_1$. This definition of unimodular time ensures that it is monotonically increasing and describes a physical ``flow of time'':
identifying it with spacetime volume to the past of $\Sigma_t$, down to a conventional time-zero $\Sigma_0$ leaf \cite{Alexander:2018djy,Magueijo:2020ntm,Magueijo:2021rpi,Magueijo:2021pvq,Jirousek:2018ago,Jirousek:2020vhy,Vikman:2021god}, we have the on-shell expression
\begin{align}\label{eq:deltatime}
    \Delta T_{\Lambda} = \int_{\Sigma_0}^{\Sigma_t} d^2x \ \sqrt{-g} =: \text{Vol}(\Sigma_t \cup \Sigma_0).
\end{align}
Simple counting of constraints shows that the theory has no new local degrees of freedom, justifying the appearance of the topological term $\partial_{\mu} \mathcal{A}^{\mu}$. There is an additional $\text{U}(1)$ gauge symmetry of the $\text{HT}_2$ action: a gauge transformation of $\mathcal{A}^{\mu}$
\begin{equation}
    \mathcal{A}^{\mu} \rightarrow \mathcal{A}^{\mu} + \chi^{\mu},
\end{equation}
with $\partial_{\mu} \chi^{\mu} = 0$ ensuring that there is one gauge degree of freedom left. This implies the invariance of the action \eqref{eq:HT2actionA}.

It is important to notice that unlike its four-dimensional cousin, the $\text{U}(1)$-reduction of HT gravity \eqref{eq:U(1)HT4action} in two dimensions is naturally emergent from the extended algebra. The Chern-Pontryagin term $F\wedge F$ in four dimensions is therefore naturally replaced in two dimensions with the first Chern class $F$. 

\subsection{Maxwell algebra and $\widehat{\text{CGHS}}$-model as flat $\text{HT}_2$ gravity}

In this section, we note that setting $\lambda = 0$ in the KSY model \eqref{eq:eqksy} yields what is known in the literature as the Cangemi–Jackiw (CJ) gravity model \cite{Cangemi:1992bj}, which can also be viewed as a Weyl rescaling of the CGHS model \cite{Callan:1992rs}. Recently, growing interest in two-dimensional flat holography called the $\reallywidehat{\text{\textcolor{black}{CGHS}}}$-model and its close connection to the complex SYK model \cite{Gonzalez:2018enk,Godet:2021cdl,Afshar:2019tvp,Afshar:2019axx,Kar:2022vqy,Kar:2022sdc,Rosso:2022tsv,Afshar:2022mkf} have reignited attention towards this model. Indeed, to study CJ theory at different temperatures, it is necessary to allow the vacuum cosmological constant $\Lambda$ to vary — precisely as achieved by the unimodular extension — which controls the Rindler inverse temperature $\beta \sim \tfrac{1}{\Lambda}$.

The CJ action is given by
\begin{equation}
    S_{\text{CJ}} = \frac{1}{2} \int d^2x \ \sqrt{-g} \bigl(\phi R - 2\Lambda\bigr),
\end{equation}
with on-shell expressions given by 
\begin{align}
    R &= 0,\\
\bigl(\nabla_{\mu} \partial_{\nu} - g_{\mu \nu} \Box\bigr)\phi &= \Lambda g_{\mu \nu}.
\end{align}
A consistent gauge construction in terms of Einstein-Cartan variables does exist for the CJ model \cite{Cangemi:1992bj,Jackiw:1992ev,Cangemi:1992ri,Grignani:1992hw,Cangemi:1993sd}. For this, we take the gauge group as the isometry group of flat two-dimensional spacetime, $\text{ISO}^{+}(1,1)$, and centrally extend it via $\text{U}(1)$. The resulting Maxwell algebra has corresponding commutation relations given by the $\lambda=0$ limit of \eqref{eq:commutationsl(2,R)}:
\begin{equation}
    \left[P_a, P_b\right] = \epsilon_{ab} I,\ \ \ \ \ [P_a, \widetilde{J}] =-\epsilon_a^{\ b} P_b.
\end{equation}

With the algebra defined as above, one can construct a gauge theory based on $\reallywidehat{\mathfrak{iso}}(1,1) \cong \mathfrak{iso}(1,1) \oplus \mathfrak{u}(1)$. Proceeding in a similar manner to the previous section, the $\lambda = 0$ curvature $\mathbf{F}$ is
\begin{equation}
    \mathbf{F} = T^a P_a + \widetilde{R} \widetilde{J} + (F + \widetilde{\Sigma}) I.
\end{equation}
The BF action \eqref{eq:BF} reduces to 
\begin{equation}
    S = \int B_a T^a +\phi \widetilde{R} -\Lambda(F+\widetilde{\Sigma}),
\end{equation}
where now $B=\phi+\Lambda$ and $\widetilde{B} = -\Lambda$. Therefore the on-shell action is given by the flat EC formulation of CJ gravity model:
\begin{align}\label{eq:eccj}
    S_{\scriptsize \reallywidehat{\text{\textcolor{black}{CGHS}}}} &= \int \phi \widetilde{R} - \Lambda(F+\widetilde{\Sigma}) \nonumber\\
    &= \frac{1}{2}\int d^2x \ \sqrt{-g} \Big[\phi R + 2 \Lambda\bigl(\epsilon^{\mu \nu} \partial_{\mu} A_{\nu} - 1\bigr) \Bigr].
\end{align}

Variation with respect to the dilaton and the metric yields the same classical dynamics as  CJ gravity, establishing that flat $\text{HT}_2$ ($\lambda=0$) and CJ gravity are on-shell equivalent theories. Moreover, the unimodular condition for flat $\text{HT}_2$ gravity remains unchanged from the $\lambda \neq 0$ $\text{HT}_2$ gravity, given by $\partial_{\mu} \mathcal{A}^{\mu} = \sqrt{-g}$.

\section{Quantum cosmology in de Sitter $\text{HT}_2$ gravity}\label{sec:qcHT2}

As an application of $\text{HT}_2$ gravity, we explore the quantum cosmology of closed universes in de Sitter space. To begin, we introduce the mini-superspace (MSS) metric of dS in KSY gravity and highlight the differences arising from the $\text{HT}_2$ model. Subsequently, we perform a canonical quantization of both theories, with a particular focus on the implications for the probability amplitudes associated with the wavefunction of the universe.

\subsection{Mini-superspace KSY gravity}

To construct the MSS KSY gravity action, the starting point is through its associated EC formulation \eqref{eq:EC HT2 action} with constant $\Lambda$:
\begin{equation}\label{eq:KSYaction}
    S_{\text{KSY}} = \int \phi( \widetilde{R}- \widetilde{\Sigma}) - \Lambda \widetilde{\Sigma},
\end{equation}
where $\Lambda$ is now just the vacuum cosmological constant and $\lambda=1$ gives the globally de Sitter spacetime geometry. Note that we are using the more general KSY model--which contains the cosmological constant term--rather than standard JT gravity, in order to introduce the HT term in the next section. In two dimensions, the MSS metric takes a rather simple form:
\begin{equation}
    ds^2 = -N^2 dt^2 + a^2 d\Sigma^2,
\end{equation}
where $N(t)$ is the lapse function (usually gauge fixed to unity) and $a(t)$ is the scale factor that often appears in cosmological models. $d\Sigma^2$ is the metric on the homogeneous and isotropic, constant-time slice, spatial hypersurface $\Sigma$. 

From the on-shell condition $R=2$, we obtain a global de Sitter geometry. Consequently, the natural choice of coordinates is global $\text{dS}_2$, where spatial hypersurfaces are closed universes described by circles $S^1$: we can identify $d\Sigma = d\theta$, where $\theta \in S^1$. The metric for a closed universe in global $\text{dS}_2$ coordinates is therefore given by
\begin{equation}
    ds^2 = -dt^2 + \cosh^2(t) d\theta^2.
\end{equation}
Geometrically, this describes a universe that is contracting at early times and expanding at later times, with a minimal length $a(0)=1$ reached at $t=0$, as the scale factor is simply $a(t) = \cosh (t)$. On the other hand, the equation of motion for the dilaton field will give the solution
\begin{equation}
    \phi(t) = \phi_0 \sinh(t) - \Lambda.
\end{equation}
Both the scale factor and dilaton expressions derived above are solutions to the Friedmann equations in a global de Sitter space.

Performing straightforward calculations, one finds that the dual area form is
\begin{equation}
    \widetilde{\Sigma} = Na \ dt \wedge d\theta,
\end{equation}
and that the dual curvature form is given by:
\begin{equation}
    \widetilde{R} = \Dot{b} \ dt \wedge d\theta,
\end{equation}
where $b = \Dot{a}/N$ is written for convenience. Since nothing in the action depends on the angular parameter $\theta$, we define the co-moving volume as $v_c = \int d\theta$. Note that in units of $v_c = 1$ (which we adopt in this paper), the scale factor $a$ and the coordinate volume $v = a v_c$ are the same. After integration by parts, the action \eqref{eq:KSYaction} in MSS reduces to
\begin{equation}
    S_{\text{KSY}} = - \int dt \ \biggl( \frac{\Dot{\phi} \Dot{a}}{N} + Na \bigl( \phi + \Lambda \bigr) \biggr),
\end{equation}
where the boundary term cancels exactly with the Gibbon-Hawking-York (GHY) term
\begin{equation}
     S_{\text{GHY}} = - \int_{\partial \mathcal{M}} d\theta \ \sqrt{|h|} \phi_{b} K = - \phi_b \frac{\Dot{a}}{N}\bigg|_{\partial \mathcal{M}},
\end{equation}
with $\phi_b$ the boundary value of the dilaton and $K$ the extrinsic curvature.

The canonical momenta are easy to read off, each associated to the dilaton and scale factor, respectively:
\begin{equation}\label{eq:KSYcanonicalmomenta}
    p_{\phi} = - \frac{\Dot{a}}{N}, \ \ \ \ \ p_a = - \frac{\Dot{\phi}}{N}.
\end{equation}
Consequently, the Hamiltonian $H$ of the MSS KSY gravity takes the form
\begin{equation}
    H = -N p_a p_{\phi} + Na \bigl( \phi + \Lambda\bigr) \equiv aN \mathcal{C},
\end{equation}
where $\mathcal{C}$ is the KSY gravity Hamiltonian constraint given by
\begin{equation}
    \mathcal{C} = - p_{a} \biggl(\frac{1}{a}\biggr) p_{\phi} + \phi + \Lambda.
\end{equation}
The action in canonical form is therefore rewritten as
\begin{equation}
    S_{\text{KSY}} = \int dt \biggl(p_{\phi} \Dot{\phi} + p_a \Dot{a} - aN \mathcal{C}\biggr).
\end{equation}
The Hamiltonian constraint equation $\mathcal{C} = 0$ is obtained as an on-shell expression when the action is varied with respect to the lapse function:
\begin{equation}
    \mathcal{C} = -p_{a} \biggl(\frac{1}{a}\biggr) p_{\phi} + \phi + \Lambda = 0.
\end{equation}
Notably, the choice of the operator ordering $p_{a} a^{-1} p_{\phi}$ in the Hamiltonian constraint is carefully chosen to ensure that the canonical quantization of the theory yields the Hartle-Hawking solution for the wavefunction of the universe.

Starting from the canonical pairs, the equal-time Poisson brackets are given by:
\begin{equation}
    \bigl\{\phi,p_{\phi}\bigr\} = \bigl\{a,p_a\bigr\} = 1.
\end{equation}
With the phase space variables promoted to linear operators acting on the KSY Hilbert space, we have the following prescription on the momentum observables:
\begin{equation}
    p_{\phi} \rightarrow \reallywidehat{p}_{\phi} = -i \frac{\partial}{\partial \phi}, \ \ \ \ \ \ \ p_{a} \rightarrow \reallywidehat{p}_{a} = -i  \frac{\partial}{\partial a}.
\end{equation}
Under this quantization scheme, the Hamiltonian constraint is transformed into the Wheeler-DeWitt (WdW) equation for MSS KSY gravity:
\begin{equation}
    \biggl[ \partial_{a} \biggl(\frac{1}{a} \partial_{\phi} \biggr) + \frac{1}{2}U'(\phi) \biggr] \psi\bigl(a,\phi,\Lambda\bigr) = 0,
\end{equation}
where we define the potential $U(\phi)=\phi^2+2\phi\Lambda$ and introduce $\psi$ as the wavefunction of the universe.

A solution to the WdW equation described above is the so-called Hartle-Hawking (HH) wavefunction \cite{Maldacena:2019cbz,Cotler:2019dcj,Cotler:2019nbi,Moitra:2022glw,Cotler:2023eza,Nanda:2023wne,Cotler:2024xzz,Anninos:2024iwf,Held:2024rmg,Buchmuller:2024ksd,Honda:2024hdr,Iizuka:2025vkl,Dey:2025osp}. We shall consider here both the expanding and contracting branches of the HH wavefunction, in the asymptotic limit with $a \phi \gg 1$. The solution then becomes a linear superposition of plane waves
\begin{equation}\label{eq:eqpsilamb}
    \psi\bigl(a,\phi,\Lambda\bigr) = \alpha_{-} \psi_{-}\bigl(a,\phi,\Lambda\bigr) + \alpha_{+} \psi_{+}\bigl(a,\phi,\Lambda\bigr),
\end{equation}
where $\psi_{\pm}\bigl(a,\phi,\Lambda\bigr) = \text{e}^{\pm ia \sqrt{U(\phi)}}$ corresponds to the contracting and expanding wavefunctions for plus and minus, respectively.

Note that from the start $\Lambda$ is a fixed constant (off-shell and on-shell). In $\text{HT}_2$ gravity, this fixed constant is promoted to an off-shell variable, but on-shell free constant. In addition, one has the freedom to consider solutions of the form $\Psi(a,\phi,\Lambda) = \mathcal{A}(\Lambda) \psi(a,\phi,\Lambda)$ for some normalizable amplitude $\mathcal{A}(\Lambda)$, which also solves the WdW equation.  An issue involving the Hartle-Hawking solution is that it is not square-integrable with respect to the DeWitt norm (a variant of the Klein-Gordon norm) \cite{Maldacena:2019cbz,Cotler:2024xzz,Held:2024rmg,Buchmuller:2024ksd}. This is partially solved by introducing normalizable amplitudes \cite{Nanda:2023wne,Dey:2025osp}. However, this requires a summation (or integration) over $\Lambda$, which is a-priori fixed in JT gravity. Furthermore, the theory fails to evolve unitarily, as pointed out in \cite{Chandrasekaran:2022cip, Witten:2023qsv, Gomez:2023wrq, Gomez:2023upk, Witten:2023xze}, an issue tied to the absence of an observer in a closed universe. In the next section, we show that both problems are resolved by transitioning from the KSY model to $\text{HT}_2$ gravity.

An alternative approach to obtain the HH wavefunctions is through a semi-classical WKB approximation, where the wavefunction is given by
\begin{equation}
    \psi(a,\phi,\Lambda) \sim \text{e}^{i S_{\text{on-shell}}},
\end{equation}
where $S_{\text{on-shell}}$ is the on-shell action for the KSY model. In this context, boundary terms play a crucial role, particularly the inclusion of the GHY term (and other possible holographic counterterms):
\begin{equation}
    S_{\partial} = S_{\text{GHY}} + S_{\text{ct}}. \nonumber
\end{equation}
The presence of a spacelike boundary at time infinity is especially significant for a holographic treatment of the theory—for instance, the Schwarzian modes residing at future infinity. However, such considerations lie beyond the scope of this paper. Accordingly, we shall disregard the boundary dynamics in our analysis and instead focus on investigating the bulk properties of MSS $\text{HT}_2$ gravity.

\subsection{Quantization of MSS $\text{HT}_2$}

As before, the Einstein-Cartan action for the two-dimensional Henneaux-Teitelboim gravity theory is given by \eqref{eq:EC HT2 action}, which in MSS takes the rather simple form:
\begin{equation}
    S_{\text{HT}_2} = - \int dt \ \biggl(\frac{\Dot{\phi}\Dot{a}}{N} - \Lambda \Dot{T}_{\Lambda}+ Na \bigl( \phi + \Lambda \bigr)\biggr).
\end{equation}
From the perspective of gauge theory, we have in the action $\Dot{T}_{\Lambda}=-\Dot{A}_{\theta}$. This is apparent in global $\text{dS}_2$ coordinates for a closed spatial slice $S^1$, where we have the unimodular time explicitly written as
\begin{equation}
    T_{\Lambda} = - \int_{S^1} A\Big|_{S^1} = - v_c A_{\theta},
\end{equation}
where in units of $v_c=1$, $T_{\Lambda}= -A_{\theta}$. Throughout this section, we shall keep the notation $T_{\Lambda}$ to denote the physical time variable.

In addition to the canonical momenta of the KSY model given in \eqref{eq:KSYcanonicalmomenta}, we gain an extra canonical momentum conjugate to the unimodular time, as expected from $\text{HT}_2$ gravity:
\begin{equation}\label{eq:clockhamiltonian}
    p_{T_{\Lambda}} = \Lambda,
\end{equation}
with equal-time Poisson bracket given by
\begin{equation}
    \bigl\{ T_{\Lambda},\Lambda \bigr\} = 1.
\end{equation}
The Hamiltonian of $\text{HT}_2$ is identical to that of KSY gravity. The action in its canonical form has the additional canonical momenta conjugate to the unimodular time:
\begin{equation}
    S_{\text{HT}_2} = \int dt \biggl(p_{\phi} \Dot{\phi} + p_a \Dot{a} + p_{T_{\Lambda}} \Dot{T}_{\Lambda} - aN \mathcal{C}\biggr).
\end{equation}
However, due to the fact that there exists a canonical momentum $\Lambda$ conjugate to $T_{\Lambda}$, one has the on shell expression 
\begin{equation}\label{eq:HT2hamiltonianconstraint}
    \mathcal{C} = \mathcal{C}_{\text{JT}} + p_{T_{\Lambda}} = \mathcal{C}_{\text{JT}} + \Lambda = 0,
\end{equation}
where $\mathcal{C}_{\text{JT}}$ is the JT gravity Hamiltonian constraint: 
\begin{equation}\label{eq:JT gravity hamiltonian}
    \mathcal{C}_{\text{JT}} = - p_{a} \biggl(\frac{1}{a}\biggr) p_{\phi} + \phi,
\end{equation}
where we use again the correct ordering of the phase space variables to reproduce the HH wavefunction solutions of the universe for $\text{HT}_2$ gravity.

Similarly to the quantization of the MSS KSY model, we shall perform a canonical quantization of $\text{HT}_2$. We demonstrate how unitary evolution and a well-defined normalizable inner product emerge within $\text{HT}_2$ gravity, thus resolving the aforementioned issues in JT gravity.

We start by first using the prescription that the canonical momenta are replaced by derivatives with respect to their conjugate pairs:
\begin{align}
    p_{\phi} \longrightarrow \reallywidehat{p}_{\phi} &= -i \frac{\partial}{\partial \phi}, \ \ \ \ p_{a} \longrightarrow  \reallywidehat{p}_{a} = -i  \frac{\partial}{\partial a},\\ 
    &p_{T_{\Lambda}} \longrightarrow  \reallywidehat{p}_{T_{\Lambda}} = -i \frac{\partial}{\partial T_{\Lambda}}.
\end{align}
From \eqref{eq:HT2hamiltonianconstraint}, the Hamiltonian constraint then becomes in the $T_{\Lambda}$ representation:
\begin{equation}
    \left[\partial_a \biggl(\frac{1}{a} \partial_{\phi}\biggr) + \phi -i \frac{\partial}{\partial T_{\Lambda}}\right] \Psi(a, \phi; T_{\Lambda}) =0.
\end{equation}
This is indeed a Schr\"{o}dinger-like equation with Hamiltonian given by $\reallywidehat{\mathcal{C}}_{\text{JT}}$:
\begin{equation}
   i \frac{\partial}{\partial T_{\Lambda}} \Psi(a, \phi; T_{\Lambda}) = \reallywidehat{\mathcal{C}}_{\text{JT}} \Psi(a, \phi; T_{\Lambda}).
\end{equation}
Notice that unlike the WdW equation in JT gravity, we now have a time-dependent wavefunction $\Psi(a, \phi; T_{\Lambda})$. To obtain this wavefunction, we need to start by determining the time-independent $\psi(a, \phi, \Lambda)$ (or $\Psi(a, \phi, \Lambda) = \mathcal{A}(\Lambda) \psi(a, \phi, \Lambda)$ since it also solves the WdW equation). This has been solved in the previous section due to the fact that the time-independent solution satisfies the WdW equation of KSY gravity. We thus need to be in the $p_{T_{\Lambda}} =\Lambda$ representation, that is writing an eigenfunction equation:
\begin{align}
    \reallywidehat{\mathcal{C}}_{\text{JT}} \Psi(a, \phi, \Lambda) &= \biggl[ \partial_{a} \biggl(\frac{1}{a} \partial_{\phi} \biggr) +\phi \biggr] \Psi(a, \phi, \Lambda) \nonumber\\
    &= - \Lambda \Psi(a, \phi, \Lambda).
\end{align}
This tells us that the energy eigenvalues for the JT gravity Hamiltonian constraint are simply given by the vacuum cosmological constant $\Lambda$.

Once again, in the context of HT gravity, $\Lambda$ is a free parameter that is constant on-shell. This allows the possibility of having a superposition of $\Lambda$-states, something that is not possible for KSY (and JT) gravity alone. Indeed, the time-dependent wavefunction can be written as a superposition of these wavefunction, multiplied by some amplitude. That is, we have solutions to the Schr\"{odinger} equation given by:
\begin{equation}\label{eq:timdependentwf}
    \Psi(a,\phi;T_{\Lambda}) = \int \frac{d\Lambda}{\sqrt{2\pi}} \ \text{e}^{i \Lambda T_{\Lambda}} \mathcal{A}\bigl(\Lambda\bigr) \psi(a,\phi, \Lambda).
\end{equation}
Here, we have used unitary evolution of the wavefunction $\Psi(a,\phi;T_{\Lambda})$, dictacted by Schr\"{o}dinger dynamics. We can now invert this formula to obtain the amplitude $\mathcal{A}(\Lambda)$ in terms of the wave function $\Psi(a,\phi;T_\Lambda)$:
\begin{equation}
    \mathcal{A}(\Lambda) =  \int da \  e^{-i\Lambda T_{\Lambda}} \psi^{\star}(a,\phi, \Lambda) \Psi(a,\phi;T_{\Lambda}).
\end{equation}

To define a physical Hilbert space, we introduce the inner product on the space of solutions to the time-dependent wavefunctions $ \Psi(a,\phi;T_{\Lambda})$ to be
\begin{equation}\label{eq:innerproduct}
    \Braket{\Psi_1}{\Psi_2} := \int d\Lambda \ \mathcal{A}^{\star}_1\bigl(\Lambda\bigr) \mathcal{A}_2\bigl(\Lambda\bigr),
\end{equation}
with norm simply given by the modulus-square of the amplitudes:
\begin{equation}
    \Braket{\Psi}{\Psi} = \int d\Lambda \ \bigl|\mathcal{A}(\Lambda)\bigr|^2.
\end{equation}
This product is automatically conserved with respect to unimodular time, i.e. unitarity is satisfied, since it is defined in terms of time-independent amplitudes. 

To construct the time-independent part of the wavefunction as a superposition of wave packets, we take the amplitude $\mathcal{A}(\Lambda)$ to be a Gaussian distribution centered at $\Lambda_0$ with standard deviation $\sigma_{T_{\Lambda}}$:
\begin{equation}\label{eq:amplitudegaussian}
    \mathcal{A}\bigl(\Lambda\bigr) = \frac{1}{(2 \pi \sigma^2_{T_{\Lambda}})^{\frac{1}{4}}} \ \exp\biggl(- \frac{(\Lambda - \Lambda_0)^2}{4 \sigma^2_{T_{\Lambda}}}\biggr).
\end{equation}
This assumption is justified by the fact that delta function amplitudes do not produce wave packet solutions to the Schr\"odinger equation, and moreover they have infinite norm, which is a non desired feature. Given that we have unitarity, it is reasonable to define physical states as any state derived from a normalized Gaussian amplitude, which satisfies $\Braket{\Psi}{\Psi}=1$.

Notably, in JT gravity, the inner product is defined by the DeWitt inner product \cite{Maldacena:2019cbz,Cotler:2024xzz,Held:2024rmg,Buchmuller:2024ksd}. This resembles the Klein-Gordon inner product, as the WdW equation can be viewed as a Klein-Gordon equation in the lightcone coordinates $(a, \phi)$. This similarity underscores the absence of standard unitary evolution for wavefunctions in JT gravity, precluding the definition of a conventional Schr\"{o}dinger inner product. However, we highlight that such a unitary evolution becomes feasible when unimodular time is introduced, enabling the definition of the physical inner product, as given by \eqref{eq:innerproduct}.
\section{Towards unitary quantum cosmology and topology change in $\text{HT}_2$}\label{sec:topologychange}
Consider the contracting and expanding branches of the HH wavefunction solution in KSY model, separately:
\begin{equation}
    \Psi_{\pm}(a,\phi,\Lambda) = \alpha_{\pm}\mathcal{A}(\Lambda) \psi_{\pm}(a,\phi,\Lambda).
\end{equation}

Then, using the normalized Gaussian amplitude \eqref{eq:amplitudegaussian} and integrating over $\Lambda$ in \eqref{eq:timdependentwf}, we find the asymptotic solution in the limit $a\phi \gg 1$:
\begin{align}
    \Psi_{\pm}(a,\phi;T_{\Lambda}) &\approx \Bigl(\frac{8 \sigma^2_{T_{\Lambda}}}{\pi}\Bigr)^{\frac{1}{4}} \text{e}^{- \sigma^2_{T_{\Lambda}}(T_{\Lambda} \pm a)^2 + i \Lambda_0 (T_{\Lambda} \pm a)} \psi_{\pm}(a,\phi),
\end{align}
where $\psi_{\pm}(a,\phi) = \text{e}^{\pm ia \phi}$ is the HH solution for MSS JT gravity. It can be further checked that this gives the correct norm as expected, i.e. from Plancherel theorem we get:
\begin{align}\label{eq:eqprob1}
    \Braket{\Psi}{\Psi} &= \int_{-\infty}^{\infty} d\Lambda \ \bigl|\mathcal{A}(\Lambda)\bigr|^2\\
    &= \int_0^{\infty} da \ \bigl|\Psi_{\pm}(a,\phi;T_{\Lambda})\bigr|^2 = 1.
\end{align}
[AE:Check the last equality using (71).]
Therefore, it is quite straightforward to see that the probability density given by $\bigl|\Psi_{\pm}(a,\phi;T_{\Lambda})\bigr|^2$, is found to follow a Gaussian distribution for a contracting and expanding closed universe, with its peak corresponding to the semi-classical WKB value.

However, we are also interested in examining linear combinations of the two branches, akin to \eqref{eq:eqpsilamb}. That is, consider the following time-dependent wavefunction solution
\begin{equation}
    \Psi(a,\phi;T_{\Lambda}) = \int \frac{d\Lambda}{\sqrt{4\pi}} \ \mathcal{A}\bigl(\Lambda\bigr) \text{e}^{i\Lambda T_{\Lambda}}\Bigl( \psi_{-} +i \psi_{+}\Bigr).
\end{equation}
Inverting the wavefunction gives an inversion formula for the amplitudes that now reads
\begin{equation}
    \mathcal{A}(\Lambda) =  \int \frac{da}{\sqrt{4\pi}} \  e^{-i\Lambda T_{\Lambda}} \Psi(a,\phi;T_{\Lambda})\Bigl(\psi_{+} - i\psi_{-} \Bigr).
\end{equation}

Consequently, we anticipate the emergence of interference terms in the probability density $\bigl|\Psi(a,\phi;T_{\Lambda})\bigr|^2$, arising from the cross-terms $\psi^{\star}_{\pm}\bigl(a,\phi\bigr)\psi_{\mp}\bigl(a,\phi\bigr)$. Indeed, we define the probability density as
\begin{eqnarray}
    {\cal P}(a,\phi,T_\Lambda) = \bigl|\Psi(a,\phi;T_{\Lambda})\bigr|^2,
    \label{eqprob}
\end{eqnarray}
where $\Psi(a,\phi;T_{\Lambda})$ is obtained by substituting (\ref{eq:eqpsilamb}) in (\ref{eq:timdependentwf}).
\begin{widetext}

\begin{figure}
    \centering
    \hspace*{-0.60cm}
    \includegraphics[scale=0.60]{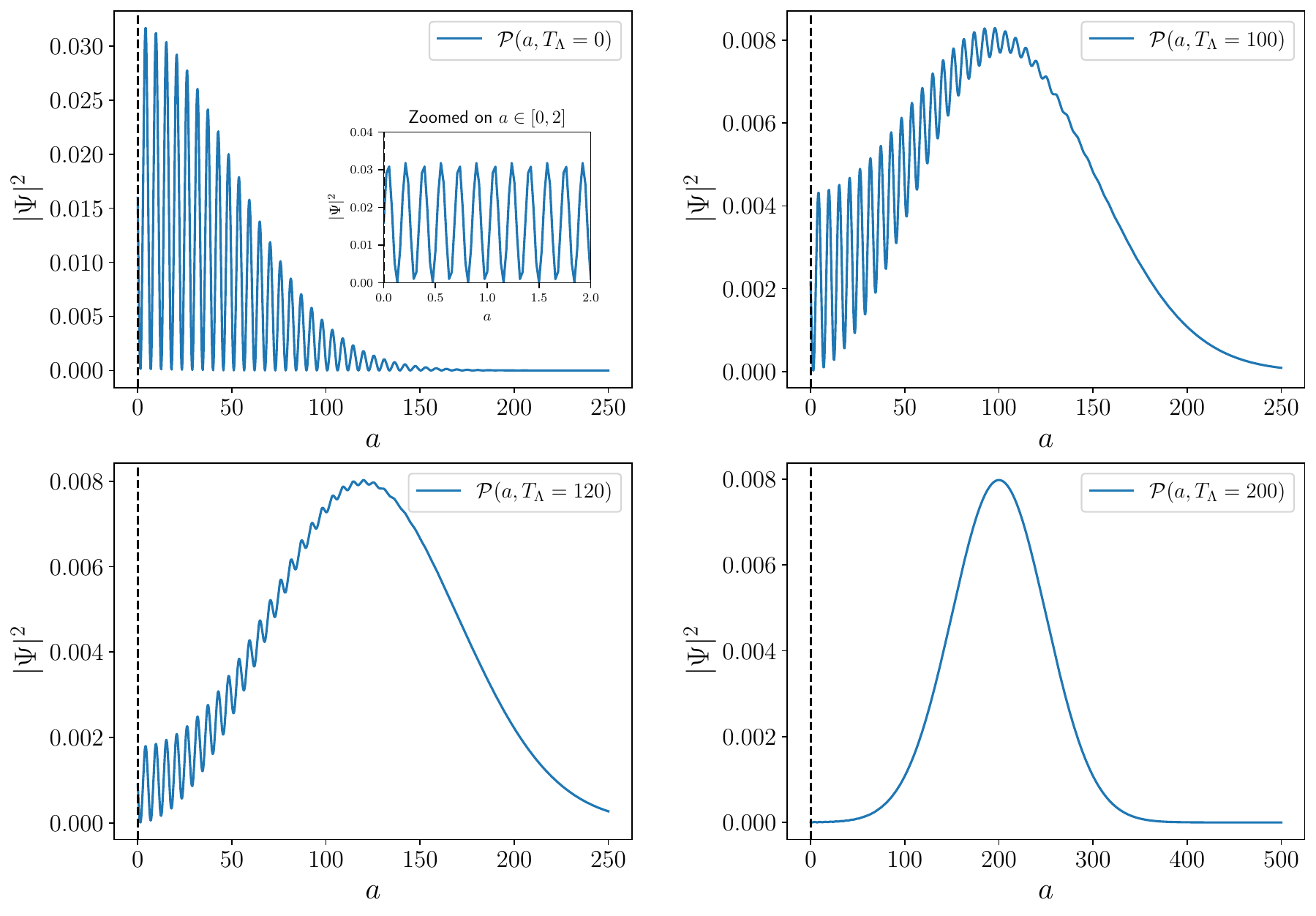}
    \caption{\textcolor{black}{Probability ${\cal P}(a,\phi,T_\Lambda)$, according to Eq.~(\ref{eqprob}), for times $T_{\Lambda}=0,$~$100,$~$120$ and 200, with $\sigma_T=0.01$ and $\phi=100$. For $|T_\Lambda|\sim \sigma_T$ the probability has many oscillations inside the envelope, but these disappear as $|T_\Lambda|$ increases, as illustrated. We see that for increasing $|T_\Lambda|\gg \sigma_T$, the width of the probability decreases, that is the uncertainty in $a$ decreases.}} 
    \label{figapsiProb}
\end{figure}

\end{widetext}
This is plotted in figure \ref{figapsiProb}.
The interference between the two quantum states becomes noticeable when $|T_\Lambda|\sim \sigma_{T_{\Lambda}}$; however, at later times, as $|T_\Lambda|\gg \sigma_{T_{\Lambda}}$, the states separate and evolve into independent, sharply peaked semiclassical WKB states that continue to narrow as unimodular time progresses. Thus, significant physics occurs around $|T_\Lambda|\sim \sigma_{T_{\Lambda}}$ (with $T_{\Lambda}=0$ in the plots). This probability density integrates to one, as it should according to \eqref{eq:eqprob1} for all times, confirming unitarity in the theory. 
Classically, the universe exhibits a bouncing de Sitter geometry in unimodular time, as shown in figure \ref{figgeometries}. To see why, consider the global de Sitter metric with scale factor $a(t) = \cosh(t)$. Imposing the on-shell unimodular condition
\begin{equation}
    \sqrt{-g} = \Dot{\mathcal{T}^0} \ \  \Longrightarrow \ \ a(t) = \Dot{T}_{\Lambda},
\end{equation}
we find that on-shell, unimodular time is
\begin{equation}
    T_{\Lambda}(t) = \int_{\Sigma_t} dt \ a(t) = \sinh(t),  \ \ \ \ \ \forall t \in \mathbb{R}.
\end{equation}
Since $T_{\Lambda}$ explicitly depends on the coordinate time $t$, we observe that $T_{\Lambda} = 0$ at $t=0$. Thus, for vanishing unimodular time, the scale factor reaches the bouncing point $a=1$.

At early unimodular times $|T_{\Lambda}| \sim \sigma_{T_{\Lambda}}$, due to quantum interference of the contracting incident and expanding reflected waves, the classical de Sitter geometry can undergo quantum deformations. In particular, the universe may evolve towards the singular point $a=0$, enabling a transition from a circular spatial topology to a point and back again. One can interpret this as a crunching universe, followed by the creation of an expanding universe at $a=0$ with a bang. Therefore in this scenario, the universe undergoes a topology change process. 
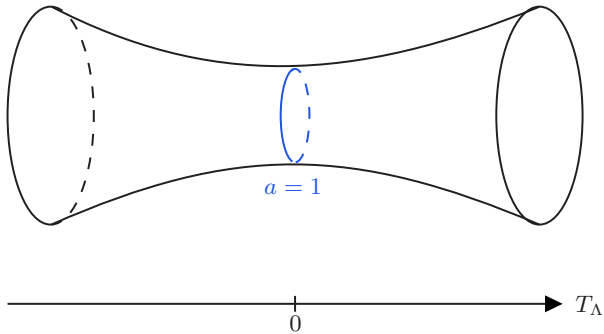
\begin{figure}\label{figgeometries} 

\vspace*{0.5cm}

\tikzset{every picture/.style={line width=0.75pt}} 

\begin{tikzpicture}[x=0.75pt,y=0.75pt,yscale=-1,xscale=1]

\draw   (288.25,50) .. controls (300.26,50) and (310,74.62) .. (310,105) .. controls (310,135.38) and (300.26,160) .. (288.25,160) .. controls (276.24,160) and (266.5,135.38) .. (266.5,105) .. controls (266.5,74.62) and (276.24,50) .. (288.25,50) -- cycle ;
\draw  [draw opacity=0] (42.11,159.99) .. controls (41.99,160) and (41.87,160) .. (41.75,160) .. controls (29.74,160) and (20,135.38) .. (20,105) .. controls (20,74.62) and (29.74,50) .. (41.75,50) .. controls (41.9,50) and (42.04,50) .. (42.19,50.01) -- (41.75,105) -- cycle ; \draw   (42.11,159.99) .. controls (41.99,160) and (41.87,160) .. (41.75,160) .. controls (29.74,160) and (20,135.38) .. (20,105) .. controls (20,74.62) and (29.74,50) .. (41.75,50) .. controls (41.9,50) and (42.04,50) .. (42.19,50.01) ;  
\draw  [draw opacity=0][dash pattern={on 4.5pt off 4.5pt}] (42.11,50.01) .. controls (53.95,50.49) and (63.5,74.92) .. (63.5,105) .. controls (63.5,135.17) and (53.9,159.66) .. (42,160) -- (41.75,105) -- cycle ; \draw  [dash pattern={on 4.5pt off 4.5pt}] (42.11,50.01) .. controls (53.95,50.49) and (63.5,74.92) .. (63.5,105) .. controls (63.5,135.17) and (53.9,159.66) .. (42,160) ;  
\draw    (288.25,160) .. controls (193.62,119.89) and (134.9,119.11) .. (42.09,159.99) ;
\draw    (288.32,50.02) .. controls (192.9,89.25) and (121.12,90.82) .. (42.16,50.01) ;
\draw  [draw opacity=0][dash pattern={on 4.5pt off 4.5pt}] (162.87,82.47) .. controls (163.54,81.79) and (164.26,81.43) .. (165,81.43) .. controls (169,81.43) and (172.25,91.98) .. (172.25,105) .. controls (172.25,118.02) and (169,128.57) .. (165,128.57) .. controls (164.52,128.57) and (164.05,128.42) .. (163.6,128.13) -- (165,105) -- cycle ; \draw  [color={rgb, 255:red, 30; green, 83; blue, 227 }  ,draw opacity=1 ][dash pattern={on 4.5pt off 4.5pt}] (162.87,82.47) .. controls (163.54,81.79) and (164.26,81.43) .. (165,81.43) .. controls (169,81.43) and (172.25,91.98) .. (172.25,105) .. controls (172.25,118.02) and (169,128.57) .. (165,128.57) .. controls (164.52,128.57) and (164.05,128.42) .. (163.6,128.13) ;  
\draw  [draw opacity=0] (163.65,128.16) .. controls (160.29,126.11) and (157.75,116.52) .. (157.75,105) .. controls (157.75,92.89) and (160.56,82.91) .. (164.18,81.58) -- (165,105) -- cycle ; \draw  [color={rgb, 255:red, 30; green, 83; blue, 227 }  ,draw opacity=1 ] (163.65,128.16) .. controls (160.29,126.11) and (157.75,116.52) .. (157.75,105) .. controls (157.75,92.89) and (160.56,82.91) .. (164.18,81.58) ;

\draw    (297,200) -- (20,200) ;
\draw [shift={(300,200)}, rotate = 180] [fill={rgb, 255:red, 0; green, 0; blue, 0 }  ][line width=0.08]  [draw opacity=0] (8.93,-4.29) -- (0,0) -- (8.93,4.29) -- cycle    ;
\draw    (165,196) -- (165,204) ;

\draw (305,194.4) node [anchor=north west][inner sep=0.75pt]    {$T_{\Lambda }$};
\draw (148,135.4) node [anchor=north west][inner sep=0.75pt]    {$\textcolor[rgb]{0.12,0.33,0.89}{a=1}$};
\draw (165.22,204.4) node [anchor=north] [inner sep=0.75pt]    {$0$};

\end{tikzpicture}
    \caption{\textcolor{black}{Schematic drawing for the on-shell unitary time evolution of global de Sitter geometry in unimodular time.}}
    \label{figgeometries}
\end{figure}

A preliminary interpretation of the results goes as follows: we extend JT gravity theory by allowing the vacuum cosmological constant $\Lambda$ to vary off-shell through the introduction of unimodular time. This modification allows the wavefunction to be in different energy eigenstates of $\Lambda$ within the quantum theory. Consequently, quantum deformations of the global de Sitter geometry at early unimodular times can be viewed as the universe occupying various $\Lambda$-dependent energy levels.

Additionally, we find that the Hartle-Hawking ``no-boundary proposal'' does not appear to be applicable in this context. The no-boundary proposal might be relevant in scenarios where the universe is exclusively expanding or contracting, with the wavefunction peaking at all unimodular times and without the presence of any quantum interference effects. However, these statements should be treated with caution, as a comprehensive path-integral formulation in this framework is needed to better illuminate these situations. Although these features warrant further investigation, a detailed discussion is beyond the scope of this paper.

\section{Page-Wootters quantization of $\text{HT}_2$ gravity}\label{sec:PW}

We have seen that two-dimensional Henneaux-Teitelboim gravity offers a consistent classical resolution to the problem of time by introducing unimodular time, which is canonically conjugate to the vacuum cosmological constant $\Lambda$. Additionally, we have developed a unitary quantum cosmological description of two-dimensional de Sitter spacetime, incorporating both the time-dependent and time-independent Hartle-Hawking wavefunctions. A key objective is to give a proper quantization scheme that aligns with the results presented in the previous section. Here, we achieve this by integrating unimodular gravity with the Page-Wootters (PW) formalism of quantum mechanics.

\subsection{Clocks, systems and the universe in the PW formalism}

In quantum mechanics, time $T$ appears as a classical parameter in the Schr\"{o}dinger equation, leading to time-dependent quantum states $\ket{\psi(T)}$, subject to unitary evolution and probability conservation. This notion of a ``physical time'' seems to be missing within quantum gravity, where quantum states are annihilated by constraint operators, indicating the absence of both unitary evolution and probability conservation, as the quantum states appear to be ``frozen in time''. It is thus of great importance to be able to give a fully quantum description of time.

The starting point of PW formalism \cite{Page:1983uc,Wootters:1984wfv,Giovannetti:2015qha,Smith:2017pwx,Hoehn:2019fsy} is to consider time as an external quantum degree of freedom by assigning to it its own ``clock'' Hilbert space $\mathcal{H}_{\text{C}}$. One can then consider clock states $\ket{T}$ in $\mathcal{H}_{\text{C}}$, defined via the time operator in its spectral decomposition
\begin{equation}
    \reallywidehat{T} = \int dT \ T  \ketbra{T}{T}.
\end{equation}
Throughout the paper, we shall assume that time is described by idealized clocks: the clock states are taken to satisfy the orthogonality condition
\begin{equation}
    \Braket{T}{T'} = \delta (T - T').
\end{equation}
With that being said, the clock Hilbert space can be viewed as the space of square-integrable function $\mathcal{H}_{\text{C}} \cong L^2(\mathbb{R})$.

Since time has an associated canonically conjugate momenta $H_{\text{C}}$ and satisfies the commutation relation on $\mathcal{H}_{\text{C}}$
\begin{equation}
    \bigl[ \reallywidehat{T},\reallywidehat{H}_{\text{C}} \bigr] = i \reallywidehat{\mathds{1}}_{\text{C}},
\end{equation}
we have the clock Hamiltonian operator $\reallywidehat{H}_{\text{C}}$ defined as
\begin{equation}
    \reallywidehat{H}_{\text{C}} = \int d\mathcal{E}_\text{C} \ \mathcal{E}_\text{C} \ketbra{\mathcal{E}_\text{C}}{\mathcal{E}_\text{C}}.
\end{equation}
Here, $\ket{\mathcal{E}_\text{C}}$ are ``clock'' energy eigenstates that lives in the clock Hilbert space, which also satisfies an orthogonality relation. Furthermore, the time and clock Hamiltonian eigenstates are Fourier duals of one another:
\begin{equation}\label{eq:fourier}
    \ket{\mathcal{E}_\text{C}} = \int \frac{dT}{\sqrt{2\pi}} \  \text{e}^{ i \mathcal{E}_\text{C} T} \ket{T}.
\end{equation}
The two eigenstates are thus related by exponentiation
\begin{equation}
    \Braket{T}{\mathcal{E}_\text{C}} = \frac{\text{e}^{i \mathcal{E}_\text{C} T}}{\sqrt{2\pi}}. 
\end{equation}
As a consequence, the clock Hamiltonian operator $\reallywidehat{H}_{\text{C}}$ in the time $T$ representation satisfies
\begin{equation}
    \reallywidehat{H}_{\text{C}} \ket{T} = -i \frac{d}{dT} \ket{T}.
\end{equation}
This will play a pivotal role in defining the Schr\"{o}dinger equation below.

To further define a quantum state of an explicit physical system, we introduce the ``system'' Hilbert space $\mathcal{H}_\text{S}$. Quantum states associated to the system are $\ket{\psi}$, and an associated Hamiltonian operator $\reallywidehat{H}_{\text{S}}$ acting on it:
\begin{equation}
    \reallywidehat{H}_{\text{S}} \ket{\psi} = E_{\text{S}} \ket{\psi},
\end{equation}
where $E_\text{S}$ is the energy (eigenvalue) of the system. The system has an associated phase space with generalized coordinates $Q=\{q\}$ and conjugate generalized momenta $P_Q=\{p_q\}$. When promoted to linear operators acting on $\mathcal{H}_{\text{S}}$, they satisfy the spectral decomposition:
\begin{align}
    \reallywidehat{Q} \ket{Q} &= Q \ket{Q},\\
    \reallywidehat{P}_Q \ket{Q} &= -i \partial_Q \ket{Q},
\end{align}
where $\ket{Q}$ is an eigenstate of the system coordinate operator. Any state $\ket{\psi} \in \mathcal{H}_{\text{S}}$ can be written as a linear superposition of $\ket{Q}$ states. The specific form of $Q$ depends on the physical system under consideration. This will be explicitly given when considering the MSS JT gravity model in the next section.

To define correlations among the clock and the system, we construct the so-called ``universe'' Hilbert space $\mathcal{H} = \mathcal{H}_\text{C} \otimes \mathcal{H}_\text{S}$. This is described by the noninteracting composite quantum system formed by the clock and system Hilbert spaces. The associated ``universe'' quantum states are then denoted by $\kket{\Psi}$. The full Hamiltonian of the universe is thus given by
\begin{equation}\label{eq:universehamiltonian}
    \reallywidehat{H} = \reallywidehat{H}_{\text{C}} \otimes \reallywidehat{\mathds{1}}_{\text{S}} + \reallywidehat{\mathds{1}}_{\text{C}} \otimes \reallywidehat{H}_{\text{S}},
\end{equation}
where the resolution of the identity on the respective subspaces are
\begin{align}
    \reallywidehat{\mathds{1}}_{\text{C}} &:= \int dT \ \ketbra{T}{T} = \int d\Lambda \ \ketbra{\Lambda}{\Lambda},\\
    \reallywidehat{\mathds{1}}_{\text{S}} &:= \int dQ \ \ketbra{Q}{Q}.
\end{align}
So far, we have given the kinematical Hilbert space. To define the physical Hilbert space of the universe, we identify physical states on $\mathcal{H}$ as those annihilated by the universe Hamiltonian:
\begin{equation}
    \reallywidehat{H} \kket{\Psi} = 0.
\end{equation}
This is akin to the Wheeler-DeWitt equation of the universe. Note that since the universe Hamiltonian annihilates physical states of the universe, they are not subject to unitary time evolution. This is what is meant by frozen (or static) quantum states. 

In order to get time-dependent quantum states subject to the standard Schr\"{o}dinger equation, we demand the system to be time-dependent. In PW formalism, this is achieved by projecting the clock to the system: the conditioned quantum states denoted $\ket{\Psi(T)} \in \mathcal{H}_{\text{S} | \text{T}}$ are then defined as follows:
\begin{equation}
    \ket{\Psi(T)} := \bra{T} \otimes \reallywidehat{\mathds{1}}_{\text{S}} \kket{\Psi}.
\end{equation}
These states are defined on a subspace of the system Hilbert space and are subject to time evolution. This can be seen by taking the Wheeler-DeWitt equation of the universe and projecting it to $\mathcal{H}_{\text{S} | \text{C}}$:
\begin{equation}
   \bigl( \bra{T} \otimes \reallywidehat{\mathds{1}}_{\text{S}} \bigr) \reallywidehat{H}\kket{\Psi} = 0 \ \ \Longrightarrow \ \  i \frac{d}{dT} \ket{\Psi(T)} = \reallywidehat{H}_{\text{S}} \ket{\Psi(T)}.
\end{equation}
We therefore get unitary time evolution of system quantum states according to
\begin{equation}
    \ket{\Psi(T)} = \text{e}^{-i T \widehat{H}_\text{S} } \ket{\Psi(0)},
\end{equation}
where $\ket{\Psi(0)} \equiv \ket{\Psi(T=0)}$ satisfies the time-independent Schr\"{odinger} equation:
\begin{equation}
   \reallywidehat{H}_\text{S} \ket{\Psi(0)} = E_{\text{S}} \ket{\Psi(0)}.
\end{equation}
It can be seen from \eqref{eq:fourier} that inverting the Fourier transform gives
\begin{equation}
    \ket{0} = \int \frac{d\mathcal{E}_\text{C}}{\sqrt{2\pi}} \ket{\mathcal{E}_\text{C}}.
\end{equation}
We can therefore write the time-independent quantum state $\ket{\Psi(0)}$ in terms of $\ket{\Psi(H_C)}$, that is:
\begin{equation}
    \ket{\Psi(0)} = \int \frac{d\mathcal{E}_\text{C}}{\sqrt{2\pi}} \ket{\Psi(\mathcal{E}_\text{C})}.
\end{equation}

We can also condition states with the clock Hamiltonian eigenkets $\ket{\mathcal{E}_\text{C}}$ and obtain the ``dual'' Schr\"{o}dinger equation:
\begin{equation}\label{eq:dualSchrodinger}
    \bigl( \bra{\mathcal{E}_\text{C}} \otimes \reallywidehat{\mathds{1}}_{\text{S}} \bigr) \reallywidehat{H}\kket{\Psi} = 0 \ \ \Longrightarrow \ \ \reallywidehat{H}_{\text{S}}\ket{\Psi(\mathcal{E}_\text{C})} = -\mathcal{E}_\text{C} \ket{\Psi(\mathcal{E}_\text{C})}.
\end{equation}
It can be read off that the energy of the system is given by $E_{\text{S}} = -\mathcal{E}_\text{C}$, in agreement with the fact that the classical Hamiltonian of the universe vanishes. 

To obtain wavefunctions of the form $\Psi(Q;T)$, we perform the bracket operation $\Braket{Q}{\Psi(T)}$. The time-dependent wavefunction solution can thus be written as a superposition of energy wavefunctions:
\begin{equation}
    \Psi(Q;T) = \int \frac{d\mathcal{E}_\text{C}}{\sqrt{2\pi}} \ \text{e}^{i \mathcal{E}_\text{C} T} \Psi(Q,\mathcal{E}_\text{C}).
\end{equation}
One can also assume that the time-independent solutions are given via a (possibly Gaussian) amplitude $\mathcal{A}(\mathcal{E}_\text{C})$, such that $\Psi(Q,\mathcal{E}_\text{C}) = \mathcal{A}(\mathcal{E}_\text{C}) \psi(Q,\mathcal{E}_\text{C})$:
\begin{equation}
    \Psi(Q;T) = \int \frac{d\mathcal{E}_\text{C}}{\sqrt{2\pi}} \ \text{e}^{i \mathcal{E}_\text{C} T} \mathcal{A}(\mathcal{E}_\text{C}) \psi(Q,\mathcal{E}_\text{C}).
\end{equation}
These are also solved by the Schr\"{odinger} equation. In this case, we obtain wave packets $\Psi(Q;T)$, whereas for $\psi(Q,\mathcal{E}_\text{C})$, these are not necessarily wave packets (for instance it can be a plane-wave solution).

[AE: Add inversion formula and define the inner product from that.]

We would like to end this section by mentioning the inner product on $\mathcal{H}$. Since we want normalizable wavefunctions on the system that are subject to probability conservation, we consider the physical inner product on the universe quantum states to be given by
\begin{equation}\label{eq:physicalinnerproduct}
    \bigl(\Psi_1|\Psi_2\bigr) := \bbra{\Psi_1} \bigl(\ketbra{T}{T} \otimes \reallywidehat{\mathds{1}}_{\text{S}}\bigr) \kket{\Psi_2}.
\end{equation}
Using the fact that the universe quantum state can be written as an entangled state of the clock and system states:
\begin{equation}\label{eq:correlatedstate}
    \kket{\Psi} = \int dT \ \ket{T} \ket{\Psi(T)},
\end{equation}
the physical inner product is written as
\begin{align}
    \bigl(\Psi_1|\Psi_2\bigr) &= \Braket{\Psi_1(T)}{\Psi_2(T)} \nonumber\\
    &= \int d\mathcal{E}_\text{C} \ \mathcal{A}_1^{\star}(\mathcal{E}_\text{C}) \mathcal{A}_2(\mathcal{E}_\text{C}).
\end{align}
Thus, the physical norm is simply 
\begin{equation}\label{eq:amplitudes}
    \bigl(\Psi|\Psi\bigr) = \int d\mathcal{E}_\text{C} \ \bigl|\mathcal{A}(\mathcal{E}_\text{C})\bigr|^2.
\end{equation}
One can also write this as an integral over $Q$ and obtain an expression for the probability density
\begin{equation}
    {\cal P}(Q,T) = \big|\Psi(Q;T)\big|^2.
\end{equation}
Indeed, from \eqref{eq:amplitudes}, if the amplitudes are normalized Gaussian distributions, then it is clear that the norm is one and thus all the probabilities add-up to one.

\subsection{PW quantization of $\text{HT}_2$ mini-superspace}

With the PW formalism established, we can now consider its application to two-dimensional Henneaux-Teitelboim gravity. For that, we first define the universe Hilbert space in terms of correlations between the clock and system subspaces. The universe being de Sitter $\text{HT}_2$ gravity in mini-superspace, the system is going to be JT gravity and the clock is given by unimodular time $T_{\Lambda}$. 

The Hilbert space of $\text{HT}_2$ is
\begin{equation}
    \mathcal{H}_{\text{HT}_2} = \mathcal{H}_{T_{\Lambda}} \otimes \mathcal{H}_{\text{JT}}.
\end{equation}
In the unimodular time subspace $\mathcal{H}_{T_{\Lambda}}$, time is an operator that satisfies $\reallywidehat{T}_{\Lambda} \ket{T_{\Lambda}} = T_{\Lambda}\ket{T_{\Lambda}}$, where unimodular time eigenstates are orthogonal, realizing idealized clocks. Similarly, the clock Hamiltonian is given by the canonically conjugate momentum of unimodular time $\reallywidehat{p}_{T_{\Lambda}} = \reallywidehat{\Lambda}$ that satisfies $\reallywidehat{\Lambda} \ket{\Lambda} = \Lambda \ket{\Lambda}$, with orthogonal $\Lambda$-eigenstates. One can therefore write \eqref{eq:correlatedstate} as
\begin{equation}\label{eq:entangledPsi}
    \kket{\Psi} = \int dT_{\Lambda} \ \ket{T_{\Lambda}} \ket{\Psi(T_{\Lambda})} = \int d\Lambda \ \ket{\Lambda} \ket{\Psi(\Lambda)},
\end{equation}
where we have the conditional states on the JT gravity Hilbert subspace:
\begin{align}
    \ket{\Psi(T_{\Lambda})} &= \bra{T_{\Lambda}} \otimes \reallywidehat{\mathds{1}}_{\text{JT}} \kket{\Psi},\\
    \ket{\Psi(\Lambda)} &= \bra{\Lambda} \otimes \reallywidehat{\mathds{1}}_{\text{JT}} \kket{\Psi}.
\end{align}

The relation \eqref{eq:entangledPsi} shows that $\kket{\Psi}$ provides a complete description of the temporal evolution of the JT gravity system by representing it in terms of correlations between
the latter and the degree of freedom of the unimodular time.

From the resolution of the identity
\begin{align}
    \reallywidehat{\mathds{1}}_{T_{\Lambda}} = \int dT_{\Lambda} \ \ketbra{T_{\Lambda}}{T_{\Lambda}} = \int d\Lambda \ \ketbra{\Lambda}{\Lambda},
\end{align}
and the fact that $\Braket{T_{\Lambda}}{\Lambda} = \tfrac{1}{\sqrt{2\pi}}\text{e}^{i \Lambda T_{\Lambda}}$, the Fourier duality expression \eqref{eq:fourier} simply becomes
\begin{equation}
    \ket{\Lambda} = \int \frac{dT_{\Lambda}}{\sqrt{2\pi}}\  \text{e}^{ i \Lambda T_{\Lambda}} \ket{T_{\Lambda}}.
\end{equation}
Indeed, at time $T_{\Lambda} = 0$, we have that
\begin{equation}\label{eq:timezerostate}
    \ket{0} = \int \frac{d\Lambda}{\sqrt{2\pi}} \ \ket{\Lambda}. 
\end{equation}
Finally, the action of $\reallywidehat{\Lambda}$ on unimodular time eigenstates is given by the derivative expression
\begin{equation}
    \reallywidehat{\Lambda} \ket{T_{\Lambda}} = -i \frac{d}{dT_{\Lambda}} \ket{T_{\Lambda}}.
\end{equation}

Going back to the Hilbert space of JT gravity, we first define the generalized coordinates as the coordinates on the mini-superspace: this is given by the scale factor $a(t)$ and the dilaton $\phi(t)$. Thus $Q=\{a,\phi\}$, with associated momenta $P_Q = \{p_a,p_{\phi}\}$. We define on $\mathcal{H}_{\text{JT}}$ the following resolution of the identity
\begin{equation}
    \reallywidehat{\mathds{1}}_{\text{JT}} := \int da \ \ketbra{a,\phi}{a,\phi},
\end{equation}
where we take constant-$\phi$ slices (corresponding to constant-coordinate time $t$ slice). Indeed, these eigenstates satisfy the orthogonality relation at equal $\phi=\phi'$:
\begin{equation}
    \Braket{a,\phi}{a',\phi} = \delta(a-a').
\end{equation}
The choice of orthogonality is motivated from the fact that we care about the size of each slice (given by the scale factor). We thus look for differently sized universes for a given constant-$\phi$ slice. 

Promoting the phase-space variables to operators acting on $\mathcal{H}_{\text{JT}}$, we have the following eigenstate equations of relevance
\begin{align}
    \reallywidehat{a} \ket{a,\phi} &= a \ket{a,\phi}, \ \ \ \ \ \ \reallywidehat{p}_a \ket{a,\phi} = -i \partial_{a} \ket{a,\phi},\\
    \reallywidehat{\phi} \ket{a,\phi} &= \phi \ket{a,\phi},\ \ \ \ \ \ \reallywidehat{p}_{\phi} \ket{a,\phi} = -i \partial_{\phi} \ket{a,\phi}.
\end{align}
States $\ket{\psi} \in \mathcal{H}_{\text{JT}}$ can be written as a superposition of eigenstates, provided the state is subject to Hamiltonian dynamics.

This is the case when we take the full Hilbert space of Henneaux-Teitelboim gravity. One can define the Hamiltonian constraint and promoting it to an operator $\reallywidehat{C}$, annihilating physical states $\kket{\Psi}$:
\begin{equation}
    \reallywidehat{\mathcal{C}} \kket{\Psi} = 0,
\end{equation}
where the Hamiltonian operator of the universe is written in terms of $\reallywidehat{\mathcal{C}}_{\text{JT}}$ (the JT gravity Hamiltonian) \eqref{eq:JT gravity hamiltonian}, and the unimodular time Hamiltonian \eqref{eq:clockhamiltonian}. For that, we take the general form of the Hamiltonian of the universe \eqref{eq:universehamiltonian} and write it as
\begin{equation}
    \reallywidehat{\mathcal{C}} = \reallywidehat{\Lambda} \otimes \reallywidehat{\mathds{1}}_{\text{JT}} + \reallywidehat{\mathds{1}}_{T_{\Lambda}} \otimes \reallywidehat{\mathcal{C}}_{\text{JT}}.
\end{equation}
The WdW equation can be conditioned to be that of Schr\"{o}dinger equation by projecting the universe states to states living in JT gravity Hilbert space:
\begin{equation}
    \bigl( \bra{T_{\Lambda}} \otimes \reallywidehat{\mathds{1}}_{\text{JT}} \bigr) \reallywidehat{\mathcal{C}}\kket{\Psi} = 0 \ \ \Longrightarrow \ \  i \frac{d}{dT_{\Lambda}} \ket{\Psi(T_{\Lambda})} = \reallywidehat{\mathcal{C}}_{\text{JT}} \ket{\Psi(T_{\Lambda})}.
\end{equation}
The quantum states $\ket{\Psi(T_{\Lambda})}$ are thus subject to unitary time evolution in unimodular time
\begin{equation}
    \ket{\Psi(T_{\Lambda})} = \text{e}^{-i T_{\Lambda} \widehat{\mathcal{C}}_{\text{JT}} } \ket{\Psi(0)}.
\end{equation}
Using \eqref{eq:timezerostate}, we have a relation between the time-dependent and independent JT gravity quantum states:
\begin{align}
    \ket{\Psi(T_{\Lambda})} &= \int \frac{d\Lambda}{\sqrt{2\pi}} \ \text{e}^{-i T_{\Lambda} \widehat{\mathcal{C}}_{\text{JT}}} \ket{\Psi(\Lambda)}\nonumber\\
    &= \int \frac{d\Lambda}{\sqrt{2\pi}} \ \text{e}^{i \Lambda T_{\Lambda}} \ket{\Psi(\Lambda)},
\end{align}
where we used the ``dual'' Schr\"{o}dinger equation \eqref{eq:dualSchrodinger}:
\begin{equation}
    \reallywidehat{\mathcal{C}}_{\text{JT}} \ket{\Psi(\Lambda)} = -\Lambda \ket{\Psi(\Lambda)} \ \ \ \ \Rightarrow \ \ \ \ E_{\text{JT}} = -\Lambda.
\end{equation}
Assuming a Gaussian state for the energy-eigenstates written as $\ket{\Psi(\Lambda)} = \mathcal{A}(\Lambda) \ket{\psi(\Lambda)}$, we have recovered a quantum state version of \eqref{eq:timdependentwf}:
\begin{equation}
    \ket{\Psi(T_{\Lambda})} = \int \frac{d\Lambda}{\sqrt{2\pi}} \ \text{e}^{i \Lambda T_{\Lambda}} \mathcal{A}(\Lambda) \ket{\psi(\Lambda)}.
\end{equation}
Note that the wavefunction expression \eqref{eq:timdependentwf} is recovered upon performing $\Braket{a,\phi}{\Psi(T_{\Lambda})} = \Psi(a,\phi;T_{\Lambda})$.

The final thing to check is whether the physical inner products match in both approaches. Indeed, using the PW inner product given in \eqref{eq:physicalinnerproduct}, we find that
\begin{align}
    \bigl(\Psi_1|\Psi_2\bigr) &= \Braket{\Psi_1(T_{\Lambda})}{\Psi_2(T_{\Lambda})} \nonumber\\
    &= \int d\Lambda \ \mathcal{A}_1^{\star}(\Lambda) \mathcal{A}_2(\Lambda) \equiv \Braket{\Psi_1}{\Psi_2}.
\end{align}
This matches the inner product defined in \eqref{eq:innerproduct}, and the norm of PW also agrees:
\begin{equation}
    \bigl(\Psi|\Psi\bigr) \equiv \Braket{\Psi_1}{\Psi_2} = \int d\Lambda \ \bigl|\mathcal{A}(\Lambda)\bigr|^2,
\end{equation}
which is unity for normalized Gaussian amplitudes.

Some comments are in order, especially regarding quantum interference effects within the formalism. Indeed, as discussed in the previous section, the WdW equation solves both for the expanding and contracting wavefunctions. That is, there is the freedom of considering their quantum superposition. In that case, we expect the emergence of interference terms at the level of probability densities. This is translated in the PW formalism by first defining the time-independent states on $\mathcal{H}_{\text{JT}}$ to be
\begin{equation}
    \ket{\psi(\Lambda)} = \frac{\ket{\psi_{-}(\Lambda)} +i  \ket{\psi_{+}(\Lambda)}}{\sqrt{2}}.
\end{equation}
Therefore, the time-dependent states are now written as
\begin{equation}
    \ket{\Psi(T_{\Lambda})} = \int \frac{d\Lambda}{\sqrt{4\pi}} \ \text{e}^{i \Lambda T_{\Lambda}} \mathcal{A}(\Lambda) \bigl(\ket{\psi_{-}(\Lambda)} + i  \ket{\psi_{+}(\Lambda)}\bigr).
\end{equation}
The inner product is known to be time-independent, hence respecting unitarity, that is
\begin{equation}
    \Braket{\Psi_1(T_{\Lambda})}{\Psi_2(T_{\Lambda})} = \Braket{\Psi_1(0)}{\Psi_2(0)}.
\end{equation}
The norm is therefore also conserved and so is the total probability. The interference terms arise at the level of the probability amplitude, which are obtained from
\begin{align}
    \Braket{\Psi(0)}{\Psi(0)} &=\int_0^{\infty} da \ \Braket{\Psi(0)}{a,\phi} \Braket{a,\phi}{\Psi(0)}\nonumber\\
    &=\int_0^{\infty} da \  \bigl|\Psi(a,\phi,\Lambda)\bigr|^2.
\end{align}
This is normalizable if the amplitudes $\mathcal{A}(\Lambda)$ are Gaussian amplitudes.

\section{Discussions} \label{sec:conclusion}

The common theme of this work is that a gauge-invariant formulation of Jackiw–Teitelboim (JT) gravity with a vacuum cosmological constant $\Lambda$ naturally emerges as a two-dimensional Henneaux–Teitelboim (HT) gravity theory. 

By gauging the isometry group of two-dimensional spacetime and introducing an abelian $\mathrm{U}(1)$ central extension, one obtains HT gravity, which shares the same on-shell dynamics--and therefore the same physical content--as JT gravity. In this framework,  $\Lambda$ is promoted to an off-shell variable that becomes fixed on-shell. Furthermore, the unimodular condition introduces a natural physical time called unimodular time that is canonically conjugate to $\Lambda$. Moreover, the flat-space analog of this formulation, known as CJ (or Weyl-rescaled CGHS) gravity, admits a similar central extension and retains the core features of ``flat'' HT gravity (or the $\reallywidehat{\text{\textcolor{black}{CGHS}}}$-model). 

With $\text{HT}_2$ gravity in hand, we illustrate its utility by performing a mini-superspace reduction in a two-dimensional quantum cosmology framework. In particular, introducing unimodular time transforms the Wheeler-DeWitt equation into a Schr\"{o}dinger-like equation, thereby ensuring unitary evolution of quantum states. Consequently, the inner product is well-defined, leading to a consistent quantum-mechanical description of two-dimensional de Sitter gravity. By solving the Schr\"{o}dinger equation, we obtain explicit normalizable and unitary time-dependent wavefunctions for the global $\text{dS}_2$ geometry, represented by expanding and contracting branches described through the unimodular Hartle–Hawking wave packets. 

Analyzing the probability density shows that, at different unimodular times $T_{\Lambda}$, the scale factor $a$ follows a probability distribution, thus providing a quantum account of the universe’s expansion and contraction. Notably, at early unimodular times, quantum interference between the expanding and contracting branches dominates. In particular, at $T_{\Lambda} = 0$, this interference persists beyond the classical region, leading to the possibility of the global de Sitter geometry to be deformed and undergo topology change at the singular point $a=0$.

The study of quantum correlations between external physical time (or ``clocks'' in terms of unimodular time) and JT gravity leads to a Page-Wootters formulation of $\text{HT}_2$ quantum gravity. Promoting the HT gravity Hamiltonian constraint to a linear operator that annihilates physical states yields the Schr\"{o}dinger equation by $T_{\Lambda}$-state conditioning. This gives unitary evolution and probability conservation of quantum states, as expected.

A natural next step is to investigate how $\text{HT}_2$ gravity behaves within a holographic framework, especially through its path-integral formulation and potential connections to random matrix theory. Studying these aspects could uncover new insights into how $\text{HT}_2$ gravity fits into broader holographic paradigms. Another promising avenue involves exploring topology change involving higher topologies and two-dimensional flat holography, given the extensive existing literature on flat holography in two dimensions \cite{Gonzalez:2018enk,Godet:2021cdl,Afshar:2019tvp,Afshar:2019axx,Kar:2022sdc,Rosso:2022tsv,Afshar:2022mkf}. Establishing concrete links between the $\text{HT}_2$ formalism and holographic approaches may not only deepen our understanding of low-dimensional quantum gravity, but also offer novel perspectives on the role of topology in holographic setups. Finally, it would be intriguing to explore whether in higher dimensions (such as three and four dimensions), HT gravity theories naturally emerge from a gauge-theoretic formulation akin to the approach developed in this work.

\section{Acknowledgments}
We would like to thank A. S. Garrido, O. Gould, F. M. Haehl, R. Isichei, J. Magueijo, I. Matthaiakakis, B. Muntz, B. Oguz, P. Saffin, and B. Tekin for helpful discussions and comments. We also thank A. Roy for assistance with the plots.

This work was supported by FCT Grant No. 2021.05694.BD (B.A.), UK Research and Innovation (UKRI) under the UK government’s Horizon Europe funding Guarantee EP/X030334/1 (A.E. funded by F.M.H.), and Royal Society Dorothy Hodgkin Fellowship (F.S.R funded by O.G.)

\bibliographystyle{apsrev4-2}
\bibliography{main}

\begin{thebibliography}{81}%
\makeatletter
\providecommand \@ifxundefined [1]{%
 \@ifx{#1\undefined}
}%
\providecommand \@ifnum [1]{%
 \ifnum #1\expandafter \@firstoftwo
 \else \expandafter \@secondoftwo
 \fi
}%
\providecommand \@ifx [1]{%
 \ifx #1\expandafter \@firstoftwo
 \else \expandafter \@secondoftwo
 \fi
}%
\providecommand \natexlab [1]{#1}%
\providecommand \enquote  [1]{``#1''}%
\providecommand \bibnamefont  [1]{#1}%
\providecommand \bibfnamefont [1]{#1}%
\providecommand \citenamefont [1]{#1}%
\providecommand \href@noop [0]{\@secondoftwo}%
\providecommand \href [0]{\begingroup \@sanitize@url \@href}%
\providecommand \@href[1]{\@@startlink{#1}\@@href}%
\providecommand \@@href[1]{\endgroup#1\@@endlink}%
\providecommand \@sanitize@url [0]{\catcode `\\12\catcode `\$12\catcode `\&12\catcode `\#12\catcode `\^12\catcode `\_12\catcode `\%12\relax}%
\providecommand \@@startlink[1]{}%
\providecommand \@@endlink[0]{}%
\providecommand \url  [0]{\begingroup\@sanitize@url \@url }%
\providecommand \@url [1]{\endgroup\@href {#1}{\urlprefix }}%
\providecommand \urlprefix  [0]{URL }%
\providecommand \Eprint [0]{\href }%
\providecommand \doibase [0]{https://doi.org/}%
\providecommand \selectlanguage [0]{\@gobble}%
\providecommand \bibinfo  [0]{\@secondoftwo}%
\providecommand \bibfield  [0]{\@secondoftwo}%
\providecommand \translation [1]{[#1]}%
\providecommand \BibitemOpen [0]{}%
\providecommand \bibitemStop [0]{}%
\providecommand \bibitemNoStop [0]{.\EOS\space}%
\providecommand \EOS [0]{\spacefactor3000\relax}%
\providecommand \BibitemShut  [1]{\csname bibitem#1\endcsname}%
\let\auto@bib@innerbib\@empty
\bibitem [{\citenamefont {Kuchar}(1991)}]{Kuchar:1991xd}%
  \BibitemOpen
  \bibfield  {author} {\bibinfo {author} {\bibfnamefont {K.~V.}\ \bibnamefont {Kuchar}},\ }\href {https://doi.org/10.1103/PhysRevD.43.3332} {\bibfield  {journal} {\bibinfo  {journal} {Phys. Rev. D}\ }\textbf {\bibinfo {volume} {43}},\ \bibinfo {pages} {3332} (\bibinfo {year} {1991})}\BibitemShut {NoStop}%
\bibitem [{\citenamefont {Bombelli}\ \emph {et~al.}(1991)\citenamefont {Bombelli}, \citenamefont {Couch},\ and\ \citenamefont {Torrence}}]{Bombelli:1991jj}%
  \BibitemOpen
  \bibfield  {author} {\bibinfo {author} {\bibfnamefont {L.}~\bibnamefont {Bombelli}}, \bibinfo {author} {\bibfnamefont {W.~E.}\ \bibnamefont {Couch}},\ and\ \bibinfo {author} {\bibfnamefont {R.~J.}\ \bibnamefont {Torrence}},\ }\href {https://doi.org/10.1103/PhysRevD.44.2589} {\bibfield  {journal} {\bibinfo  {journal} {Phys. Rev. D}\ }\textbf {\bibinfo {volume} {44}},\ \bibinfo {pages} {2589} (\bibinfo {year} {1991})}\BibitemShut {NoStop}%
\bibitem [{\citenamefont {Daughton}\ \emph {et~al.}(1993)\citenamefont {Daughton}, \citenamefont {Louko},\ and\ \citenamefont {Sorkin}}]{Daughton:1993uy}%
  \BibitemOpen
  \bibfield  {author} {\bibinfo {author} {\bibfnamefont {A.}~\bibnamefont {Daughton}}, \bibinfo {author} {\bibfnamefont {J.}~\bibnamefont {Louko}},\ and\ \bibinfo {author} {\bibfnamefont {R.~D.}\ \bibnamefont {Sorkin}},\ }in\ \href@noop {} {\emph {\bibinfo {booktitle} {{5th Canadian Conference on General Relativity and Relativistic Astrophysics (5CCGRRA)}}}}\ (\bibinfo {year} {1993})\ \Eprint {https://arxiv.org/abs/gr-qc/9305016} {arXiv:gr-qc/9305016} \BibitemShut {NoStop}%
\bibitem [{\citenamefont {Sorkin}(1994)}]{sorkin1}%
  \BibitemOpen
  \bibfield  {author} {\bibinfo {author} {\bibfnamefont {R.~D.}\ \bibnamefont {Sorkin}},\ }\href {https://doi.org/10.1007/BF00670514} {\bibfield  {journal} {\bibinfo  {journal} {Int J Theor Phys}\ }\textbf {\bibinfo {volume} {33}},\ \bibinfo {pages} {523–534} (\bibinfo {year} {1994})}\BibitemShut {NoStop}%
\bibitem [{\citenamefont {Daughton}\ \emph {et~al.}(1998)\citenamefont {Daughton}, \citenamefont {Louko},\ and\ \citenamefont {Sorkin}}]{Daughton:1998aa}%
  \BibitemOpen
  \bibfield  {author} {\bibinfo {author} {\bibfnamefont {A.}~\bibnamefont {Daughton}}, \bibinfo {author} {\bibfnamefont {J.}~\bibnamefont {Louko}},\ and\ \bibinfo {author} {\bibfnamefont {R.~D.}\ \bibnamefont {Sorkin}},\ }\href {https://doi.org/10.1103/PhysRevD.58.084008} {\bibfield  {journal} {\bibinfo  {journal} {Phys. Rev. D}\ }\textbf {\bibinfo {volume} {58}},\ \bibinfo {pages} {084008} (\bibinfo {year} {1998})},\ \Eprint {https://arxiv.org/abs/gr-qc/9805101} {arXiv:gr-qc/9805101} \BibitemShut {NoStop}%
\bibitem [{\citenamefont {{Einstein}}(1952)}]{Einstein-unimod}%
  \BibitemOpen
  \bibfield  {author} {\bibinfo {author} {\bibfnamefont {A.}~\bibnamefont {{Einstein}}},\ }\bibinfo {title} {Do gravitational fields play an essential part in the structure of the elementary particles of matter?},\ in\ \href {https://ui.adsabs.harvard.edu/abs/1952prel.book..189E} {\emph {\bibinfo {booktitle} {The Principle of Relativity}}},\ \bibinfo {editor} {edited by\ \bibinfo {editor} {\bibfnamefont {A.}~\bibnamefont {Einstein}}}\ (\bibinfo  {publisher} {Dover Books on Physics},\ \bibinfo {year} {1952})\ p.\ \bibinfo {pages} {189–198}\BibitemShut {NoStop}%
\bibitem [{\citenamefont {Weinberg}(1989)}]{Weinberg:1988cp}%
  \BibitemOpen
  \bibfield  {author} {\bibinfo {author} {\bibfnamefont {S.}~\bibnamefont {Weinberg}},\ }\href {https://doi.org/10.1103/RevModPhys.61.1} {\bibfield  {journal} {\bibinfo  {journal} {Rev. Mod. Phys.}\ }\textbf {\bibinfo {volume} {61}},\ \bibinfo {pages} {1} (\bibinfo {year} {1989})}\BibitemShut {NoStop}%
\bibitem [{\citenamefont {Henneaux}\ and\ \citenamefont {Teitelboim}(1989)}]{Henneaux:1989zc}%
  \BibitemOpen
  \bibfield  {author} {\bibinfo {author} {\bibfnamefont {M.}~\bibnamefont {Henneaux}}\ and\ \bibinfo {author} {\bibfnamefont {C.}~\bibnamefont {Teitelboim}},\ }\href {https://doi.org/10.1016/0370-2693(89)91251-3} {\bibfield  {journal} {\bibinfo  {journal} {Phys. Lett. B}\ }\textbf {\bibinfo {volume} {222}},\ \bibinfo {pages} {195} (\bibinfo {year} {1989})}\BibitemShut {NoStop}%
\bibitem [{\citenamefont {Unruh}(1989)}]{unimod1}%
  \BibitemOpen
  \bibfield  {author} {\bibinfo {author} {\bibfnamefont {W.~G.}\ \bibnamefont {Unruh}},\ }\href {https://journals.aps.org/prd/abstract/10.1103/PhysRevD.40.1048} {\bibfield  {journal} {\bibinfo  {journal} {Phys. Rev. D}\ }\textbf {\bibinfo {volume} {40}},\ \bibinfo {pages} {1048} (\bibinfo {year} {1989})}\BibitemShut {NoStop}%
\bibitem [{\citenamefont {Smolin}(2009)}]{Smolin:2009ti}%
  \BibitemOpen
  \bibfield  {author} {\bibinfo {author} {\bibfnamefont {L.}~\bibnamefont {Smolin}},\ }\href {https://doi.org/10.1103/PhysRevD.80.084003} {\bibfield  {journal} {\bibinfo  {journal} {Phys. Rev. D}\ }\textbf {\bibinfo {volume} {80}},\ \bibinfo {pages} {084003} (\bibinfo {year} {2009})},\ \Eprint {https://arxiv.org/abs/0904.4841} {arXiv:0904.4841 [hep-th]} \BibitemShut {NoStop}%
\bibitem [{\citenamefont {Smolin}(2011)}]{Smolin:2010iq}%
  \BibitemOpen
  \bibfield  {author} {\bibinfo {author} {\bibfnamefont {L.}~\bibnamefont {Smolin}},\ }\href {https://doi.org/10.1103/PhysRevD.84.044047} {\bibfield  {journal} {\bibinfo  {journal} {Phys. Rev. D}\ }\textbf {\bibinfo {volume} {84}},\ \bibinfo {pages} {044047} (\bibinfo {year} {2011})},\ \Eprint {https://arxiv.org/abs/1008.1759} {arXiv:1008.1759 [hep-th]} \BibitemShut {NoStop}%
\bibitem [{\citenamefont {Kaloper}\ and\ \citenamefont {Padilla}(2014)}]{Kaloper:2013zca}%
  \BibitemOpen
  \bibfield  {author} {\bibinfo {author} {\bibfnamefont {N.}~\bibnamefont {Kaloper}}\ and\ \bibinfo {author} {\bibfnamefont {A.}~\bibnamefont {Padilla}},\ }\href {https://doi.org/10.1103/PhysRevLett.112.091304} {\bibfield  {journal} {\bibinfo  {journal} {Phys. Rev. Lett.}\ }\textbf {\bibinfo {volume} {112}},\ \bibinfo {pages} {091304} (\bibinfo {year} {2014})},\ \Eprint {https://arxiv.org/abs/1309.6562} {arXiv:1309.6562 [hep-th]} \BibitemShut {NoStop}%
\bibitem [{\citenamefont {Fern\'andez~Crist\'obal}(2014)}]{FernandezCristobal:2014jca}%
  \BibitemOpen
  \bibfield  {author} {\bibinfo {author} {\bibfnamefont {J.~M.}\ \bibnamefont {Fern\'andez~Crist\'obal}},\ }\href {https://doi.org/10.1016/j.aop.2014.07.046} {\bibfield  {journal} {\bibinfo  {journal} {Annals Phys.}\ }\textbf {\bibinfo {volume} {350}},\ \bibinfo {pages} {441} (\bibinfo {year} {2014})}\BibitemShut {NoStop}%
\bibitem [{\citenamefont {Padilla}\ and\ \citenamefont {Saltas}(2015)}]{Padilla:2014yea}%
  \BibitemOpen
  \bibfield  {author} {\bibinfo {author} {\bibfnamefont {A.}~\bibnamefont {Padilla}}\ and\ \bibinfo {author} {\bibfnamefont {I.~D.}\ \bibnamefont {Saltas}},\ }\href {https://doi.org/10.1140/epjc/s10052-015-3767-0} {\bibfield  {journal} {\bibinfo  {journal} {Eur. Phys. J. C}\ }\textbf {\bibinfo {volume} {75}},\ \bibinfo {pages} {561} (\bibinfo {year} {2015})},\ \Eprint {https://arxiv.org/abs/1409.3573} {arXiv:1409.3573 [gr-qc]} \BibitemShut {NoStop}%
\bibitem [{\citenamefont {Bufalo}\ \emph {et~al.}(2015)\citenamefont {Bufalo}, \citenamefont {Oksanen},\ and\ \citenamefont {Tureanu}}]{Bufalo:2015wda}%
  \BibitemOpen
  \bibfield  {author} {\bibinfo {author} {\bibfnamefont {R.}~\bibnamefont {Bufalo}}, \bibinfo {author} {\bibfnamefont {M.}~\bibnamefont {Oksanen}},\ and\ \bibinfo {author} {\bibfnamefont {A.}~\bibnamefont {Tureanu}},\ }\href {https://doi.org/10.1140/epjc/s10052-015-3683-3} {\bibfield  {journal} {\bibinfo  {journal} {Eur. Phys. J. C}\ }\textbf {\bibinfo {volume} {75}},\ \bibinfo {pages} {477} (\bibinfo {year} {2015})},\ \Eprint {https://arxiv.org/abs/1505.04978} {arXiv:1505.04978 [hep-th]} \BibitemShut {NoStop}%
\bibitem [{\citenamefont {Kaloper}\ \emph {et~al.}(2016)\citenamefont {Kaloper}, \citenamefont {Padilla}, \citenamefont {Stefanyszyn},\ and\ \citenamefont {Zahariade}}]{pad1}%
  \BibitemOpen
  \bibfield  {author} {\bibinfo {author} {\bibfnamefont {N.}~\bibnamefont {Kaloper}}, \bibinfo {author} {\bibfnamefont {A.}~\bibnamefont {Padilla}}, \bibinfo {author} {\bibfnamefont {D.}~\bibnamefont {Stefanyszyn}},\ and\ \bibinfo {author} {\bibfnamefont {G.}~\bibnamefont {Zahariade}},\ }\href {https://link.aps.org/doi/10.1103/PhysRevLett.116.051302} {\bibfield  {journal} {\bibinfo  {journal} {Phys. Rev. Lett.}\ }\textbf {\bibinfo {volume} {116}},\ \bibinfo {pages} {051302} (\bibinfo {year} {2016})}\BibitemShut {NoStop}%
\bibitem [{\citenamefont {Percacci}(2018)}]{Percacci:2017fsy}%
  \BibitemOpen
  \bibfield  {author} {\bibinfo {author} {\bibfnamefont {R.}~\bibnamefont {Percacci}},\ }\href {https://doi.org/10.1007/s10701-018-0189-5} {\bibfield  {journal} {\bibinfo  {journal} {Found. Phys.}\ }\textbf {\bibinfo {volume} {48}},\ \bibinfo {pages} {1364} (\bibinfo {year} {2018})},\ \Eprint {https://arxiv.org/abs/1712.09903} {arXiv:1712.09903 [gr-qc]} \BibitemShut {NoStop}%
\bibitem [{\citenamefont {Lombriser}(2019)}]{lombriser2019cosmological}%
  \BibitemOpen
  \bibfield  {author} {\bibinfo {author} {\bibfnamefont {L.}~\bibnamefont {Lombriser}},\ }\href {https://doi.org/10.1016/j.physletb.2019.134804} {\bibfield  {journal} {\bibinfo  {journal} {Phys. Lett. B}\ }\textbf {\bibinfo {volume} {797}},\ \bibinfo {pages} {134804} (\bibinfo {year} {2019})}\BibitemShut {NoStop}%
\bibitem [{\citenamefont {Carballo-Rubio}\ \emph {et~al.}(2022)\citenamefont {Carballo-Rubio}, \citenamefont {Garay},\ and\ \citenamefont {Garc\'\i{}a-Moreno}}]{Carballo-Rubio:2022ofy}%
  \BibitemOpen
  \bibfield  {author} {\bibinfo {author} {\bibfnamefont {R.}~\bibnamefont {Carballo-Rubio}}, \bibinfo {author} {\bibfnamefont {L.~J.}\ \bibnamefont {Garay}},\ and\ \bibinfo {author} {\bibfnamefont {G.}~\bibnamefont {Garc\'\i{}a-Moreno}},\ }\href {https://doi.org/10.1088/1361-6382/aca386} {\bibfield  {journal} {\bibinfo  {journal} {Class. Quant. Grav.}\ }\textbf {\bibinfo {volume} {39}},\ \bibinfo {pages} {243001} (\bibinfo {year} {2022})},\ \Eprint {https://arxiv.org/abs/2207.08499} {arXiv:2207.08499 [gr-qc]} \BibitemShut {NoStop}%
\bibitem [{\citenamefont {Alexander}\ \emph {et~al.}(2019)\citenamefont {Alexander}, \citenamefont {Magueijo},\ and\ \citenamefont {Smolin}}]{Alexander:2018djy}%
  \BibitemOpen
  \bibfield  {author} {\bibinfo {author} {\bibfnamefont {S.}~\bibnamefont {Alexander}}, \bibinfo {author} {\bibfnamefont {J.}~\bibnamefont {Magueijo}},\ and\ \bibinfo {author} {\bibfnamefont {L.}~\bibnamefont {Smolin}},\ }\href {https://doi.org/10.3390/sym11091130} {\bibfield  {journal} {\bibinfo  {journal} {Symmetry}\ }\textbf {\bibinfo {volume} {11}},\ \bibinfo {pages} {1130} (\bibinfo {year} {2019})},\ \Eprint {https://arxiv.org/abs/1807.01381} {arXiv:1807.01381 [gr-qc]} \BibitemShut {NoStop}%
\bibitem [{\citenamefont {Magueijo}\ \emph {et~al.}(2020)\citenamefont {Magueijo}, \citenamefont {Zlosnik},\ and\ \citenamefont {Speziale}}]{Magueijo:2020ntm}%
  \BibitemOpen
  \bibfield  {author} {\bibinfo {author} {\bibfnamefont {J.}~\bibnamefont {Magueijo}}, \bibinfo {author} {\bibfnamefont {T.}~\bibnamefont {Zlosnik}},\ and\ \bibinfo {author} {\bibfnamefont {S.}~\bibnamefont {Speziale}},\ }\href {https://doi.org/10.1103/PhysRevD.102.064006} {\bibfield  {journal} {\bibinfo  {journal} {Phys. Rev. D}\ }\textbf {\bibinfo {volume} {102}},\ \bibinfo {pages} {064006} (\bibinfo {year} {2020})},\ \Eprint {https://arxiv.org/abs/2006.05766} {arXiv:2006.05766 [gr-qc]} \BibitemShut {NoStop}%
\bibitem [{\citenamefont {Magueijo}(2021)}]{Magueijo:2021rpi}%
  \BibitemOpen
  \bibfield  {author} {\bibinfo {author} {\bibfnamefont {J.}~\bibnamefont {Magueijo}},\ }\href {https://doi.org/10.1016/j.physletb.2021.136487} {\bibfield  {journal} {\bibinfo  {journal} {Phys. Lett. B}\ }\textbf {\bibinfo {volume} {820}},\ \bibinfo {pages} {136487} (\bibinfo {year} {2021})},\ \Eprint {https://arxiv.org/abs/2104.11529} {arXiv:2104.11529 [gr-qc]} \BibitemShut {NoStop}%
\bibitem [{\citenamefont {Magueijo}(2022)}]{Magueijo:2021pvq}%
  \BibitemOpen
  \bibfield  {author} {\bibinfo {author} {\bibfnamefont {J.}~\bibnamefont {Magueijo}},\ }\href {https://doi.org/10.1103/PhysRevD.106.084021} {\bibfield  {journal} {\bibinfo  {journal} {Phys. Rev. D}\ }\textbf {\bibinfo {volume} {106}},\ \bibinfo {pages} {084021} (\bibinfo {year} {2022})},\ \Eprint {https://arxiv.org/abs/2110.05920} {arXiv:2110.05920 [gr-qc]} \BibitemShut {NoStop}%
\bibitem [{\citenamefont {Jirou\v{s}ek}\ and\ \citenamefont {Vikman}(2019)}]{Jirousek:2018ago}%
  \BibitemOpen
  \bibfield  {author} {\bibinfo {author} {\bibfnamefont {P.}~\bibnamefont {Jirou\v{s}ek}}\ and\ \bibinfo {author} {\bibfnamefont {A.}~\bibnamefont {Vikman}},\ }\href {https://doi.org/10.1088/1475-7516/2019/04/004} {\bibfield  {journal} {\bibinfo  {journal} {JCAP}\ }\textbf {\bibinfo {volume} {04}},\ \bibinfo {pages} {004}},\ \Eprint {https://arxiv.org/abs/1811.09547} {arXiv:1811.09547 [gr-qc]} \BibitemShut {NoStop}%
\bibitem [{\citenamefont {Jirou\v{s}ek}\ \emph {et~al.}(2021)\citenamefont {Jirou\v{s}ek}, \citenamefont {Shimada}, \citenamefont {Vikman},\ and\ \citenamefont {Yamaguchi}}]{Jirousek:2020vhy}%
  \BibitemOpen
  \bibfield  {author} {\bibinfo {author} {\bibfnamefont {P.}~\bibnamefont {Jirou\v{s}ek}}, \bibinfo {author} {\bibfnamefont {K.}~\bibnamefont {Shimada}}, \bibinfo {author} {\bibfnamefont {A.}~\bibnamefont {Vikman}},\ and\ \bibinfo {author} {\bibfnamefont {M.}~\bibnamefont {Yamaguchi}},\ }\href {https://iopscience.iop.org/article/10.1088/1475-7516/2021/04/028} {\bibfield  {journal} {\bibinfo  {journal} {JCAP}\ }\textbf {\bibinfo {volume} {04}},\ \bibinfo {pages} {028}}\BibitemShut {NoStop}%
\bibitem [{\citenamefont {Vikman}(2021)}]{Vikman:2021god}%
  \BibitemOpen
  \bibfield  {author} {\bibinfo {author} {\bibfnamefont {A.}~\bibnamefont {Vikman}},\ }in\ \href@noop {} {\emph {\bibinfo {booktitle} {{55th Rencontres de Moriond on Gravitation}}}}\ (\bibinfo {year} {2021})\ \Eprint {https://arxiv.org/abs/2107.09601} {arXiv:2107.09601 [gr-qc]} \BibitemShut {NoStop}%
\bibitem [{\citenamefont {Alexandre}\ and\ \citenamefont {Magueijo}(2023)}]{Alexandre:2022npo}%
  \BibitemOpen
  \bibfield  {author} {\bibinfo {author} {\bibfnamefont {B.}~\bibnamefont {Alexandre}}\ and\ \bibinfo {author} {\bibfnamefont {J.~a.}\ \bibnamefont {Magueijo}},\ }\href {https://doi.org/10.1103/PhysRevD.107.063501} {\bibfield  {journal} {\bibinfo  {journal} {Phys. Rev. D}\ }\textbf {\bibinfo {volume} {107}},\ \bibinfo {pages} {063501} (\bibinfo {year} {2023})},\ \Eprint {https://arxiv.org/abs/2210.02179} {arXiv:2210.02179 [hep-th]} \BibitemShut {NoStop}%
\bibitem [{\citenamefont {Alexandre}\ \emph {et~al.}(2023)\citenamefont {Alexandre}, \citenamefont {Isichei},\ and\ \citenamefont {Magueijo}}]{Alexandre:2023ozf}%
  \BibitemOpen
  \bibfield  {author} {\bibinfo {author} {\bibfnamefont {B.}~\bibnamefont {Alexandre}}, \bibinfo {author} {\bibfnamefont {R.}~\bibnamefont {Isichei}},\ and\ \bibinfo {author} {\bibfnamefont {J.~a.}\ \bibnamefont {Magueijo}},\ }\href {https://doi.org/10.1103/PhysRevD.108.023526} {\bibfield  {journal} {\bibinfo  {journal} {Phys. Rev. D}\ }\textbf {\bibinfo {volume} {108}},\ \bibinfo {pages} {023526} (\bibinfo {year} {2023})},\ \Eprint {https://arxiv.org/abs/2304.00666} {arXiv:2304.00666 [hep-th]} \BibitemShut {NoStop}%
\bibitem [{\citenamefont {Magueijo}\ and\ \citenamefont {Isichei}(2023)}]{PathInt}%
  \BibitemOpen
  \bibfield  {author} {\bibinfo {author} {\bibfnamefont {J.}~\bibnamefont {Magueijo}}\ and\ \bibinfo {author} {\bibfnamefont {R.}~\bibnamefont {Isichei}},\ }\href {https://doi.org/10.1103/PhysRevD.107.023526} {\bibfield  {journal} {\bibinfo  {journal} {Phys. Rev. D}\ }\textbf {\bibinfo {volume} {107}},\ \bibinfo {pages} {023526} (\bibinfo {year} {2023})}\BibitemShut {NoStop}%
\bibitem [{\citenamefont {Chandrasekaran}\ \emph {et~al.}(2023)\citenamefont {Chandrasekaran}, \citenamefont {Longo}, \citenamefont {Penington},\ and\ \citenamefont {Witten}}]{Chandrasekaran:2022cip}%
  \BibitemOpen
  \bibfield  {author} {\bibinfo {author} {\bibfnamefont {V.}~\bibnamefont {Chandrasekaran}}, \bibinfo {author} {\bibfnamefont {R.}~\bibnamefont {Longo}}, \bibinfo {author} {\bibfnamefont {G.}~\bibnamefont {Penington}},\ and\ \bibinfo {author} {\bibfnamefont {E.}~\bibnamefont {Witten}},\ }\href {https://doi.org/10.1007/JHEP02(2023)082} {\bibfield  {journal} {\bibinfo  {journal} {JHEP}\ }\textbf {\bibinfo {volume} {02}},\ \bibinfo {pages} {082}},\ \Eprint {https://arxiv.org/abs/2206.10780} {arXiv:2206.10780 [hep-th]} \BibitemShut {NoStop}%
\bibitem [{\citenamefont {Witten}(2024{\natexlab{a}})}]{Witten:2023qsv}%
  \BibitemOpen
  \bibfield  {author} {\bibinfo {author} {\bibfnamefont {E.}~\bibnamefont {Witten}},\ }\href {https://doi.org/10.1090/pspum/107/01954} {\bibfield  {journal} {\bibinfo  {journal} {Proc. Symp. Pure Math.}\ }\textbf {\bibinfo {volume} {107}},\ \bibinfo {pages} {247} (\bibinfo {year} {2024}{\natexlab{a}})},\ \Eprint {https://arxiv.org/abs/2303.02837} {arXiv:2303.02837 [hep-th]} \BibitemShut {NoStop}%
\bibitem [{\citenamefont {Gomez}(2023{\natexlab{a}})}]{Gomez:2023wrq}%
  \BibitemOpen
  \bibfield  {author} {\bibinfo {author} {\bibfnamefont {C.}~\bibnamefont {Gomez}},\ }\href@noop {} {\  (\bibinfo {year} {2023}{\natexlab{a}})},\ \Eprint {https://arxiv.org/abs/2302.14747} {arXiv:2302.14747 [hep-th]} \BibitemShut {NoStop}%
\bibitem [{\citenamefont {Gomez}(2023{\natexlab{b}})}]{Gomez:2023upk}%
  \BibitemOpen
  \bibfield  {author} {\bibinfo {author} {\bibfnamefont {C.}~\bibnamefont {Gomez}},\ }\href@noop {} {\  (\bibinfo {year} {2023}{\natexlab{b}})},\ \Eprint {https://arxiv.org/abs/2304.11845} {arXiv:2304.11845 [hep-th]} \BibitemShut {NoStop}%
\bibitem [{\citenamefont {Witten}(2024{\natexlab{b}})}]{Witten:2023xze}%
  \BibitemOpen
  \bibfield  {author} {\bibinfo {author} {\bibfnamefont {E.}~\bibnamefont {Witten}},\ }\href {https://doi.org/10.1007/JHEP03(2024)077} {\bibfield  {journal} {\bibinfo  {journal} {JHEP}\ }\textbf {\bibinfo {volume} {03}},\ \bibinfo {pages} {077}},\ \Eprint {https://arxiv.org/abs/2308.03663} {arXiv:2308.03663 [hep-th]} \BibitemShut {NoStop}%
\bibitem [{\citenamefont {Page}\ and\ \citenamefont {Wootters}(1983)}]{Page:1983uc}%
  \BibitemOpen
  \bibfield  {author} {\bibinfo {author} {\bibfnamefont {D.~N.}\ \bibnamefont {Page}}\ and\ \bibinfo {author} {\bibfnamefont {W.~K.}\ \bibnamefont {Wootters}},\ }\href {https://doi.org/10.1103/PhysRevD.27.2885} {\bibfield  {journal} {\bibinfo  {journal} {Phys. Rev. D}\ }\textbf {\bibinfo {volume} {27}},\ \bibinfo {pages} {2885} (\bibinfo {year} {1983})}\BibitemShut {NoStop}%
\bibitem [{\citenamefont {Wootters}(1984)}]{Wootters:1984wfv}%
  \BibitemOpen
  \bibfield  {author} {\bibinfo {author} {\bibfnamefont {W.~K.}\ \bibnamefont {Wootters}},\ }\href {https://doi.org/10.1007/BF02214098} {\bibfield  {journal} {\bibinfo  {journal} {Int. J. Theor. Phys.}\ }\textbf {\bibinfo {volume} {23}},\ \bibinfo {pages} {701} (\bibinfo {year} {1984})}\BibitemShut {NoStop}%
\bibitem [{\citenamefont {Giovannetti}\ \emph {et~al.}(2015)\citenamefont {Giovannetti}, \citenamefont {Lloyd},\ and\ \citenamefont {Maccone}}]{Giovannetti:2015qha}%
  \BibitemOpen
  \bibfield  {author} {\bibinfo {author} {\bibfnamefont {V.}~\bibnamefont {Giovannetti}}, \bibinfo {author} {\bibfnamefont {S.}~\bibnamefont {Lloyd}},\ and\ \bibinfo {author} {\bibfnamefont {L.}~\bibnamefont {Maccone}},\ }\href {https://doi.org/10.1103/PhysRevD.92.045033} {\bibfield  {journal} {\bibinfo  {journal} {Phys. Rev. D}\ }\textbf {\bibinfo {volume} {92}},\ \bibinfo {pages} {045033} (\bibinfo {year} {2015})},\ \Eprint {https://arxiv.org/abs/1504.04215} {arXiv:1504.04215 [quant-ph]} \BibitemShut {NoStop}%
\bibitem [{\citenamefont {Smith}\ and\ \citenamefont {Ahmadi}(2019)}]{Smith:2017pwx}%
  \BibitemOpen
  \bibfield  {author} {\bibinfo {author} {\bibfnamefont {A.~R.~H.}\ \bibnamefont {Smith}}\ and\ \bibinfo {author} {\bibfnamefont {M.}~\bibnamefont {Ahmadi}},\ }\href {https://doi.org/10.22331/q-2019-07-08-160} {\bibfield  {journal} {\bibinfo  {journal} {Quantum}\ }\textbf {\bibinfo {volume} {3}},\ \bibinfo {pages} {160} (\bibinfo {year} {2019})},\ \Eprint {https://arxiv.org/abs/1712.00081} {arXiv:1712.00081 [quant-ph]} \BibitemShut {NoStop}%
\bibitem [{\citenamefont {Hohn}\ \emph {et~al.}(2021)\citenamefont {Hohn}, \citenamefont {Smith},\ and\ \citenamefont {Lock}}]{Hoehn:2019fsy}%
  \BibitemOpen
  \bibfield  {author} {\bibinfo {author} {\bibfnamefont {P.~A.}\ \bibnamefont {Hohn}}, \bibinfo {author} {\bibfnamefont {A.~R.~H.}\ \bibnamefont {Smith}},\ and\ \bibinfo {author} {\bibfnamefont {M.~P.~E.}\ \bibnamefont {Lock}},\ }\href {https://doi.org/10.1103/PhysRevD.104.066001} {\bibfield  {journal} {\bibinfo  {journal} {Phys. Rev. D}\ }\textbf {\bibinfo {volume} {104}},\ \bibinfo {pages} {066001} (\bibinfo {year} {2021})},\ \Eprint {https://arxiv.org/abs/1912.00033} {arXiv:1912.00033 [quant-ph]} \BibitemShut {NoStop}%
\bibitem [{\citenamefont {De~Vuyst}\ \emph {et~al.}(2024{\natexlab{a}})\citenamefont {De~Vuyst}, \citenamefont {Eccles}, \citenamefont {Hohn},\ and\ \citenamefont {Kirklin}}]{DeVuyst:2024pop}%
  \BibitemOpen
  \bibfield  {author} {\bibinfo {author} {\bibfnamefont {J.}~\bibnamefont {De~Vuyst}}, \bibinfo {author} {\bibfnamefont {S.}~\bibnamefont {Eccles}}, \bibinfo {author} {\bibfnamefont {P.~A.}\ \bibnamefont {Hohn}},\ and\ \bibinfo {author} {\bibfnamefont {J.}~\bibnamefont {Kirklin}},\ }\href@noop {} {\  (\bibinfo {year} {2024}{\natexlab{a}})},\ \Eprint {https://arxiv.org/abs/2405.00114} {arXiv:2405.00114 [hep-th]} \BibitemShut {NoStop}%
\bibitem [{\citenamefont {De~Vuyst}\ \emph {et~al.}(2024{\natexlab{b}})\citenamefont {De~Vuyst}, \citenamefont {Eccles}, \citenamefont {Hohn},\ and\ \citenamefont {Kirklin}}]{DeVuyst:2024uvd}%
  \BibitemOpen
  \bibfield  {author} {\bibinfo {author} {\bibfnamefont {J.}~\bibnamefont {De~Vuyst}}, \bibinfo {author} {\bibfnamefont {S.}~\bibnamefont {Eccles}}, \bibinfo {author} {\bibfnamefont {P.~A.}\ \bibnamefont {Hohn}},\ and\ \bibinfo {author} {\bibfnamefont {J.}~\bibnamefont {Kirklin}},\ }\href@noop {} {\  (\bibinfo {year} {2024}{\natexlab{b}})},\ \Eprint {https://arxiv.org/abs/2412.15502} {arXiv:2412.15502 [hep-th]} \BibitemShut {NoStop}%
\bibitem [{\citenamefont {Cangemi}\ and\ \citenamefont {Jackiw}(1992)}]{Cangemi:1992bj}%
  \BibitemOpen
  \bibfield  {author} {\bibinfo {author} {\bibfnamefont {D.}~\bibnamefont {Cangemi}}\ and\ \bibinfo {author} {\bibfnamefont {R.}~\bibnamefont {Jackiw}},\ }\href {https://doi.org/10.1103/PhysRevLett.69.233} {\bibfield  {journal} {\bibinfo  {journal} {Phys. Rev. Lett.}\ }\textbf {\bibinfo {volume} {69}},\ \bibinfo {pages} {233} (\bibinfo {year} {1992})},\ \Eprint {https://arxiv.org/abs/hep-th/9203056} {arXiv:hep-th/9203056} \BibitemShut {NoStop}%
\bibitem [{\citenamefont {Jackiw}(1992)}]{Jackiw:1992ev}%
  \BibitemOpen
  \bibfield  {author} {\bibinfo {author} {\bibfnamefont {R.}~\bibnamefont {Jackiw}},\ }in\ \href@noop {} {\emph {\bibinfo {booktitle} {{19th International Colloquium on Group Theoretical Methods in Physics}}}}\ (\bibinfo {year} {1992})\BibitemShut {NoStop}%
\bibitem [{\citenamefont {Cangemi}(1992)}]{Cangemi:1992ri}%
  \BibitemOpen
  \bibfield  {author} {\bibinfo {author} {\bibfnamefont {D.}~\bibnamefont {Cangemi}},\ }\href {https://doi.org/10.1016/0370-2693(92)91259-C} {\bibfield  {journal} {\bibinfo  {journal} {Phys. Lett. B}\ }\textbf {\bibinfo {volume} {297}},\ \bibinfo {pages} {261} (\bibinfo {year} {1992})},\ \Eprint {https://arxiv.org/abs/gr-qc/9207004} {arXiv:gr-qc/9207004} \BibitemShut {NoStop}%
\bibitem [{\citenamefont {Grignani}\ and\ \citenamefont {Nardelli}(1994)}]{Grignani:1992hw}%
  \BibitemOpen
  \bibfield  {author} {\bibinfo {author} {\bibfnamefont {G.}~\bibnamefont {Grignani}}\ and\ \bibinfo {author} {\bibfnamefont {G.}~\bibnamefont {Nardelli}},\ }\href {https://doi.org/10.1016/0550-3213(94)90505-3} {\bibfield  {journal} {\bibinfo  {journal} {Nucl. Phys. B}\ }\textbf {\bibinfo {volume} {412}},\ \bibinfo {pages} {320} (\bibinfo {year} {1994})},\ \Eprint {https://arxiv.org/abs/gr-qc/9209013} {arXiv:gr-qc/9209013} \BibitemShut {NoStop}%
\bibitem [{\citenamefont {Cangemi}\ and\ \citenamefont {Jackiw}(1993{\natexlab{a}})}]{Cangemi:1992up}%
  \BibitemOpen
  \bibfield  {author} {\bibinfo {author} {\bibfnamefont {D.}~\bibnamefont {Cangemi}}\ and\ \bibinfo {author} {\bibfnamefont {R.}~\bibnamefont {Jackiw}},\ }\href {https://doi.org/10.1016/0370-2693(93)90878-L} {\bibfield  {journal} {\bibinfo  {journal} {Phys. Lett. B}\ }\textbf {\bibinfo {volume} {299}},\ \bibinfo {pages} {24} (\bibinfo {year} {1993}{\natexlab{a}})},\ \Eprint {https://arxiv.org/abs/hep-th/9210036} {arXiv:hep-th/9210036} \BibitemShut {NoStop}%
\bibitem [{\citenamefont {Kim}\ \emph {et~al.}(1993{\natexlab{a}})\citenamefont {Kim}, \citenamefont {Soh},\ and\ \citenamefont {Yee}}]{Kim:1992fb}%
  \BibitemOpen
  \bibfield  {author} {\bibinfo {author} {\bibfnamefont {S.~K.}\ \bibnamefont {Kim}}, \bibinfo {author} {\bibfnamefont {K.~S.}\ \bibnamefont {Soh}},\ and\ \bibinfo {author} {\bibfnamefont {J.-H.}\ \bibnamefont {Yee}},\ }\href {https://doi.org/10.1103/PhysRevD.47.4433} {\bibfield  {journal} {\bibinfo  {journal} {Phys. Rev. D}\ }\textbf {\bibinfo {volume} {47}},\ \bibinfo {pages} {4433} (\bibinfo {year} {1993}{\natexlab{a}})}\BibitemShut {NoStop}%
\bibitem [{\citenamefont {Kim}\ \emph {et~al.}(1993{\natexlab{b}})\citenamefont {Kim}, \citenamefont {Soh},\ and\ \citenamefont {Yee}}]{Kim:1992ht}%
  \BibitemOpen
  \bibfield  {author} {\bibinfo {author} {\bibfnamefont {S.~K.}\ \bibnamefont {Kim}}, \bibinfo {author} {\bibfnamefont {K.~S.}\ \bibnamefont {Soh}},\ and\ \bibinfo {author} {\bibfnamefont {J.-H.}\ \bibnamefont {Yee}},\ }\href {https://doi.org/10.1016/0370-2693(93)90357-N} {\bibfield  {journal} {\bibinfo  {journal} {Phys. Lett. B}\ }\textbf {\bibinfo {volume} {300}},\ \bibinfo {pages} {223} (\bibinfo {year} {1993}{\natexlab{b}})}\BibitemShut {NoStop}%
\bibitem [{\citenamefont {Cangemi}\ and\ \citenamefont {Jackiw}(1993{\natexlab{b}})}]{Cangemi:1993sd}%
  \BibitemOpen
  \bibfield  {author} {\bibinfo {author} {\bibfnamefont {D.}~\bibnamefont {Cangemi}}\ and\ \bibinfo {author} {\bibfnamefont {R.}~\bibnamefont {Jackiw}},\ }\href {https://doi.org/10.1006/aphy.1993.1058} {\bibfield  {journal} {\bibinfo  {journal} {Annals Phys.}\ }\textbf {\bibinfo {volume} {225}},\ \bibinfo {pages} {229} (\bibinfo {year} {1993}{\natexlab{b}})},\ \Eprint {https://arxiv.org/abs/hep-th/9302026} {arXiv:hep-th/9302026} \BibitemShut {NoStop}%
\bibitem [{\citenamefont {Cangemi}\ and\ \citenamefont {Dunne}(1993)}]{Cangemi:1993bb}%
  \BibitemOpen
  \bibfield  {author} {\bibinfo {author} {\bibfnamefont {D.}~\bibnamefont {Cangemi}}\ and\ \bibinfo {author} {\bibfnamefont {G.~V.}\ \bibnamefont {Dunne}},\ }\href {https://doi.org/10.1103/PhysRevD.48.5721} {\bibfield  {journal} {\bibinfo  {journal} {Phys. Rev. D}\ }\textbf {\bibinfo {volume} {48}},\ \bibinfo {pages} {5721} (\bibinfo {year} {1993})},\ \Eprint {https://arxiv.org/abs/hep-th/9308021} {arXiv:hep-th/9308021} \BibitemShut {NoStop}%
\bibitem [{\citenamefont {Jackiw}(1993)}]{Jackiw:1993gf}%
  \BibitemOpen
  \bibfield  {author} {\bibinfo {author} {\bibfnamefont {R.}~\bibnamefont {Jackiw}},\ }in\ \href@noop {} {\emph {\bibinfo {booktitle} {{International Conference on Interface Between Physics and Mathematics (IPM 93) (Part 1: Plenary Session, 6-10 Sep 1993: Part 2: Tutorial Lectures, 13-17 Sep 1993)}}}}\ (\bibinfo {year} {1993})\ pp.\ \bibinfo {pages} {116--129},\ \Eprint {https://arxiv.org/abs/hep-th/9309082} {arXiv:hep-th/9309082} \BibitemShut {NoStop}%
\bibitem [{\citenamefont {Hammer}\ \emph {et~al.}(2020)\citenamefont {Hammer}, \citenamefont {Jirousek},\ and\ \citenamefont {Vikman}}]{Hammer:2020dqp}%
  \BibitemOpen
  \bibfield  {author} {\bibinfo {author} {\bibfnamefont {K.}~\bibnamefont {Hammer}}, \bibinfo {author} {\bibfnamefont {P.}~\bibnamefont {Jirousek}},\ and\ \bibinfo {author} {\bibfnamefont {A.}~\bibnamefont {Vikman}},\ }\href@noop {} {\  (\bibinfo {year} {2020})},\ \Eprint {https://arxiv.org/abs/2001.03169} {arXiv:2001.03169 [gr-qc]} \BibitemShut {NoStop}%
\bibitem [{\citenamefont {Etkin}\ \emph {et~al.}(2024)\citenamefont {Etkin}, \citenamefont {Magueijo},\ and\ \citenamefont {Rassouli}}]{Etkin:2023amf}%
  \BibitemOpen
  \bibfield  {author} {\bibinfo {author} {\bibfnamefont {A.}~\bibnamefont {Etkin}}, \bibinfo {author} {\bibfnamefont {J.}~\bibnamefont {Magueijo}},\ and\ \bibinfo {author} {\bibfnamefont {F.-S.}\ \bibnamefont {Rassouli}},\ }\href {https://doi.org/10.1016/j.physletb.2024.138810} {\bibfield  {journal} {\bibinfo  {journal} {Phys. Lett. B}\ }\textbf {\bibinfo {volume} {855}},\ \bibinfo {pages} {138810} (\bibinfo {year} {2024})},\ \Eprint {https://arxiv.org/abs/2311.11160} {arXiv:2311.11160 [hep-th]} \BibitemShut {NoStop}%
\bibitem [{\citenamefont {Gonz\'alez}\ \emph {et~al.}(2018)\citenamefont {Gonz\'alez}, \citenamefont {Grumiller},\ and\ \citenamefont {Salzer}}]{Gonzalez:2018enk}%
  \BibitemOpen
  \bibfield  {author} {\bibinfo {author} {\bibfnamefont {H.~A.}\ \bibnamefont {Gonz\'alez}}, \bibinfo {author} {\bibfnamefont {D.}~\bibnamefont {Grumiller}},\ and\ \bibinfo {author} {\bibfnamefont {J.}~\bibnamefont {Salzer}},\ }\href {https://doi.org/10.1007/JHEP05(2018)083} {\bibfield  {journal} {\bibinfo  {journal} {JHEP}\ }\textbf {\bibinfo {volume} {05}},\ \bibinfo {pages} {083}},\ \Eprint {https://arxiv.org/abs/1802.01562} {arXiv:1802.01562 [hep-th]} \BibitemShut {NoStop}%
\bibitem [{\citenamefont {Afshar}(2020)}]{Afshar:2019tvp}%
  \BibitemOpen
  \bibfield  {author} {\bibinfo {author} {\bibfnamefont {H.~R.}\ \bibnamefont {Afshar}},\ }\href {https://doi.org/10.1007/JHEP02(2020)126} {\bibfield  {journal} {\bibinfo  {journal} {JHEP}\ }\textbf {\bibinfo {volume} {02}},\ \bibinfo {pages} {126}},\ \Eprint {https://arxiv.org/abs/1908.08089} {arXiv:1908.08089 [hep-th]} \BibitemShut {NoStop}%
\bibitem [{\citenamefont {Afshar}\ \emph {et~al.}(2020)\citenamefont {Afshar}, \citenamefont {Gonz\'alez}, \citenamefont {Grumiller},\ and\ \citenamefont {Vassilevich}}]{Afshar:2019axx}%
  \BibitemOpen
  \bibfield  {author} {\bibinfo {author} {\bibfnamefont {H.}~\bibnamefont {Afshar}}, \bibinfo {author} {\bibfnamefont {H.~A.}\ \bibnamefont {Gonz\'alez}}, \bibinfo {author} {\bibfnamefont {D.}~\bibnamefont {Grumiller}},\ and\ \bibinfo {author} {\bibfnamefont {D.}~\bibnamefont {Vassilevich}},\ }\href {https://doi.org/10.1103/PhysRevD.101.086024} {\bibfield  {journal} {\bibinfo  {journal} {Phys. Rev. D}\ }\textbf {\bibinfo {volume} {101}},\ \bibinfo {pages} {086024} (\bibinfo {year} {2020})},\ \Eprint {https://arxiv.org/abs/1911.05739} {arXiv:1911.05739 [hep-th]} \BibitemShut {NoStop}%
\bibitem [{\citenamefont {Godet}\ and\ \citenamefont {Marteau}(2021)}]{Godet:2021cdl}%
  \BibitemOpen
  \bibfield  {author} {\bibinfo {author} {\bibfnamefont {V.}~\bibnamefont {Godet}}\ and\ \bibinfo {author} {\bibfnamefont {C.}~\bibnamefont {Marteau}},\ }\href {https://doi.org/10.1007/JHEP07(2021)138} {\bibfield  {journal} {\bibinfo  {journal} {JHEP}\ }\textbf {\bibinfo {volume} {07}},\ \bibinfo {pages} {138}},\ \Eprint {https://arxiv.org/abs/2103.13422} {arXiv:2103.13422 [hep-th]} \BibitemShut {NoStop}%
\bibitem [{\citenamefont {Afshar}\ and\ \citenamefont {Oblak}(2022)}]{Afshar:2021qvi}%
  \BibitemOpen
  \bibfield  {author} {\bibinfo {author} {\bibfnamefont {H.}~\bibnamefont {Afshar}}\ and\ \bibinfo {author} {\bibfnamefont {B.}~\bibnamefont {Oblak}},\ }\href {https://doi.org/10.1007/JHEP11(2022)172} {\bibfield  {journal} {\bibinfo  {journal} {JHEP}\ }\textbf {\bibinfo {volume} {11}},\ \bibinfo {pages} {172}},\ \Eprint {https://arxiv.org/abs/2112.14609} {arXiv:2112.14609 [hep-th]} \BibitemShut {NoStop}%
\bibitem [{\citenamefont {Kar}\ \emph {et~al.}(2023)\citenamefont {Kar}, \citenamefont {Lamprou}, \citenamefont {Marteau},\ and\ \citenamefont {Rosso}}]{Kar:2022sdc}%
  \BibitemOpen
  \bibfield  {author} {\bibinfo {author} {\bibfnamefont {A.}~\bibnamefont {Kar}}, \bibinfo {author} {\bibfnamefont {L.}~\bibnamefont {Lamprou}}, \bibinfo {author} {\bibfnamefont {C.}~\bibnamefont {Marteau}},\ and\ \bibinfo {author} {\bibfnamefont {F.}~\bibnamefont {Rosso}},\ }\href {https://doi.org/10.1007/JHEP03(2023)249} {\bibfield  {journal} {\bibinfo  {journal} {JHEP}\ }\textbf {\bibinfo {volume} {03}},\ \bibinfo {pages} {249}},\ \Eprint {https://arxiv.org/abs/2208.05974} {arXiv:2208.05974 [hep-th]} \BibitemShut {NoStop}%
\bibitem [{\citenamefont {Rosso}(2023)}]{Rosso:2022tsv}%
  \BibitemOpen
  \bibfield  {author} {\bibinfo {author} {\bibfnamefont {F.}~\bibnamefont {Rosso}},\ }\href {https://doi.org/10.1007/JHEP02(2023)037} {\bibfield  {journal} {\bibinfo  {journal} {JHEP}\ }\textbf {\bibinfo {volume} {02}},\ \bibinfo {pages} {037}},\ \Eprint {https://arxiv.org/abs/2209.14372} {arXiv:2209.14372 [hep-th]} \BibitemShut {NoStop}%
\bibitem [{\citenamefont {Afshar}\ and\ \citenamefont {Aghamir}(2023)}]{Afshar:2022mkf}%
  \BibitemOpen
  \bibfield  {author} {\bibinfo {author} {\bibfnamefont {H.}~\bibnamefont {Afshar}}\ and\ \bibinfo {author} {\bibfnamefont {N.}~\bibnamefont {Aghamir}},\ }\href {https://doi.org/10.1007/JHEP03(2023)009} {\bibfield  {journal} {\bibinfo  {journal} {JHEP}\ }\textbf {\bibinfo {volume} {03}},\ \bibinfo {pages} {009}},\ \Eprint {https://arxiv.org/abs/2211.00612} {arXiv:2211.00612 [hep-th]} \BibitemShut {NoStop}%
\bibitem [{\citenamefont {Callan}\ \emph {et~al.}(1992)\citenamefont {Callan}, \citenamefont {Giddings}, \citenamefont {Harvey},\ and\ \citenamefont {Strominger}}]{Callan:1992rs}%
  \BibitemOpen
  \bibfield  {author} {\bibinfo {author} {\bibfnamefont {C.~G.}\ \bibnamefont {Callan}, \bibfnamefont {Jr.}}, \bibinfo {author} {\bibfnamefont {S.~B.}\ \bibnamefont {Giddings}}, \bibinfo {author} {\bibfnamefont {J.~A.}\ \bibnamefont {Harvey}},\ and\ \bibinfo {author} {\bibfnamefont {A.}~\bibnamefont {Strominger}},\ }\href {https://doi.org/10.1103/PhysRevD.45.R1005} {\bibfield  {journal} {\bibinfo  {journal} {Phys. Rev. D}\ }\textbf {\bibinfo {volume} {45}},\ \bibinfo {pages} {R1005} (\bibinfo {year} {1992})},\ \Eprint {https://arxiv.org/abs/hep-th/9111056} {arXiv:hep-th/9111056} \BibitemShut {NoStop}%
\bibitem [{\citenamefont {Godet}\ and\ \citenamefont {Marteau}(2020)}]{Godet:2020xpk}%
  \BibitemOpen
  \bibfield  {author} {\bibinfo {author} {\bibfnamefont {V.}~\bibnamefont {Godet}}\ and\ \bibinfo {author} {\bibfnamefont {C.}~\bibnamefont {Marteau}},\ }\href {https://doi.org/10.1007/JHEP12(2020)020} {\bibfield  {journal} {\bibinfo  {journal} {JHEP}\ }\textbf {\bibinfo {volume} {12}},\ \bibinfo {pages} {020}},\ \Eprint {https://arxiv.org/abs/2005.08999} {arXiv:2005.08999 [hep-th]} \BibitemShut {NoStop}%
\bibitem [{\citenamefont {Teitelboim}(1983)}]{Teitelboim:1983ux}%
  \BibitemOpen
  \bibfield  {author} {\bibinfo {author} {\bibfnamefont {C.}~\bibnamefont {Teitelboim}},\ }\href {https://doi.org/10.1016/0370-2693(83)90012-6} {\bibfield  {journal} {\bibinfo  {journal} {Phys. Lett. B}\ }\textbf {\bibinfo {volume} {126}},\ \bibinfo {pages} {41} (\bibinfo {year} {1983})}\BibitemShut {NoStop}%
\bibitem [{\citenamefont {Jackiw}(1985)}]{Jackiw:1984je}%
  \BibitemOpen
  \bibfield  {author} {\bibinfo {author} {\bibfnamefont {R.}~\bibnamefont {Jackiw}},\ }\href {https://doi.org/10.1016/0550-3213(85)90448-1} {\bibfield  {journal} {\bibinfo  {journal} {Nucl. Phys. B}\ }\textbf {\bibinfo {volume} {252}},\ \bibinfo {pages} {343} (\bibinfo {year} {1985})}\BibitemShut {NoStop}%
\bibitem [{\citenamefont {Isler}\ and\ \citenamefont {Trugenberger}(1989)}]{Isler:1989hq}%
  \BibitemOpen
  \bibfield  {author} {\bibinfo {author} {\bibfnamefont {K.}~\bibnamefont {Isler}}\ and\ \bibinfo {author} {\bibfnamefont {C.~A.}\ \bibnamefont {Trugenberger}},\ }\href {https://doi.org/10.1103/PhysRevLett.63.834} {\bibfield  {journal} {\bibinfo  {journal} {Phys. Rev. Lett.}\ }\textbf {\bibinfo {volume} {63}},\ \bibinfo {pages} {834} (\bibinfo {year} {1989})}\BibitemShut {NoStop}%
\bibitem [{\citenamefont {Chamseddine}\ and\ \citenamefont {Wyler}(1989)}]{Chamseddine:1989yz}%
  \BibitemOpen
  \bibfield  {author} {\bibinfo {author} {\bibfnamefont {A.~H.}\ \bibnamefont {Chamseddine}}\ and\ \bibinfo {author} {\bibfnamefont {D.}~\bibnamefont {Wyler}},\ }\href {https://doi.org/10.1016/0370-2693(89)90528-5} {\bibfield  {journal} {\bibinfo  {journal} {Phys. Lett. B}\ }\textbf {\bibinfo {volume} {228}},\ \bibinfo {pages} {75} (\bibinfo {year} {1989})}\BibitemShut {NoStop}%
\bibitem [{\citenamefont {Kar}\ \emph {et~al.}(2022)\citenamefont {Kar}, \citenamefont {Lamprou}, \citenamefont {Marteau},\ and\ \citenamefont {Rosso}}]{Kar:2022vqy}%
  \BibitemOpen
  \bibfield  {author} {\bibinfo {author} {\bibfnamefont {A.}~\bibnamefont {Kar}}, \bibinfo {author} {\bibfnamefont {L.}~\bibnamefont {Lamprou}}, \bibinfo {author} {\bibfnamefont {C.}~\bibnamefont {Marteau}},\ and\ \bibinfo {author} {\bibfnamefont {F.}~\bibnamefont {Rosso}},\ }\href {https://doi.org/10.1103/PhysRevLett.129.201601} {\bibfield  {journal} {\bibinfo  {journal} {Phys. Rev. Lett.}\ }\textbf {\bibinfo {volume} {129}},\ \bibinfo {pages} {201601} (\bibinfo {year} {2022})},\ \Eprint {https://arxiv.org/abs/2205.02240} {arXiv:2205.02240 [hep-th]} \BibitemShut {NoStop}%
\bibitem [{\citenamefont {Maldacena}\ \emph {et~al.}(2021)\citenamefont {Maldacena}, \citenamefont {Turiaci},\ and\ \citenamefont {Yang}}]{Maldacena:2019cbz}%
  \BibitemOpen
  \bibfield  {author} {\bibinfo {author} {\bibfnamefont {J.}~\bibnamefont {Maldacena}}, \bibinfo {author} {\bibfnamefont {G.~J.}\ \bibnamefont {Turiaci}},\ and\ \bibinfo {author} {\bibfnamefont {Z.}~\bibnamefont {Yang}},\ }\href {https://doi.org/10.1007/JHEP01(2021)139} {\bibfield  {journal} {\bibinfo  {journal} {JHEP}\ }\textbf {\bibinfo {volume} {01}},\ \bibinfo {pages} {139}},\ \Eprint {https://arxiv.org/abs/1904.01911} {arXiv:1904.01911 [hep-th]} \BibitemShut {NoStop}%
\bibitem [{\citenamefont {Cotler}\ and\ \citenamefont {Jensen}(2021)}]{Cotler:2019dcj}%
  \BibitemOpen
  \bibfield  {author} {\bibinfo {author} {\bibfnamefont {J.}~\bibnamefont {Cotler}}\ and\ \bibinfo {author} {\bibfnamefont {K.}~\bibnamefont {Jensen}},\ }\href {https://doi.org/10.1007/JHEP12(2021)089} {\bibfield  {journal} {\bibinfo  {journal} {JHEP}\ }\textbf {\bibinfo {volume} {12}},\ \bibinfo {pages} {089}},\ \Eprint {https://arxiv.org/abs/1911.12358} {arXiv:1911.12358 [hep-th]} \BibitemShut {NoStop}%
\bibitem [{\citenamefont {Cotler}\ \emph {et~al.}(2020)\citenamefont {Cotler}, \citenamefont {Jensen},\ and\ \citenamefont {Maloney}}]{Cotler:2019nbi}%
  \BibitemOpen
  \bibfield  {author} {\bibinfo {author} {\bibfnamefont {J.}~\bibnamefont {Cotler}}, \bibinfo {author} {\bibfnamefont {K.}~\bibnamefont {Jensen}},\ and\ \bibinfo {author} {\bibfnamefont {A.}~\bibnamefont {Maloney}},\ }\href {https://doi.org/10.1007/JHEP06(2020)048} {\bibfield  {journal} {\bibinfo  {journal} {JHEP}\ }\textbf {\bibinfo {volume} {06}},\ \bibinfo {pages} {048}},\ \Eprint {https://arxiv.org/abs/1905.03780} {arXiv:1905.03780 [hep-th]} \BibitemShut {NoStop}%
\bibitem [{\citenamefont {Moitra}\ \emph {et~al.}(2022)\citenamefont {Moitra}, \citenamefont {Sake},\ and\ \citenamefont {Trivedi}}]{Moitra:2022glw}%
  \BibitemOpen
  \bibfield  {author} {\bibinfo {author} {\bibfnamefont {U.}~\bibnamefont {Moitra}}, \bibinfo {author} {\bibfnamefont {S.~K.}\ \bibnamefont {Sake}},\ and\ \bibinfo {author} {\bibfnamefont {S.~P.}\ \bibnamefont {Trivedi}},\ }\href {https://doi.org/10.1007/JHEP06(2022)138} {\bibfield  {journal} {\bibinfo  {journal} {JHEP}\ }\textbf {\bibinfo {volume} {06}},\ \bibinfo {pages} {138}},\ \Eprint {https://arxiv.org/abs/2202.03130} {arXiv:2202.03130 [hep-th]} \BibitemShut {NoStop}%
\bibitem [{\citenamefont {Cotler}\ and\ \citenamefont {Jensen}(2023)}]{Cotler:2023eza}%
  \BibitemOpen
  \bibfield  {author} {\bibinfo {author} {\bibfnamefont {J.}~\bibnamefont {Cotler}}\ and\ \bibinfo {author} {\bibfnamefont {K.}~\bibnamefont {Jensen}},\ }\href {https://doi.org/10.1103/PhysRevLett.131.211601} {\bibfield  {journal} {\bibinfo  {journal} {Phys. Rev. Lett.}\ }\textbf {\bibinfo {volume} {131}},\ \bibinfo {pages} {211601} (\bibinfo {year} {2023})},\ \Eprint {https://arxiv.org/abs/2302.06603} {arXiv:2302.06603 [hep-th]} \BibitemShut {NoStop}%
\bibitem [{\citenamefont {Nanda}\ \emph {et~al.}(2024)\citenamefont {Nanda}, \citenamefont {Sake},\ and\ \citenamefont {Trivedi}}]{Nanda:2023wne}%
  \BibitemOpen
  \bibfield  {author} {\bibinfo {author} {\bibfnamefont {K.~K.}\ \bibnamefont {Nanda}}, \bibinfo {author} {\bibfnamefont {S.~K.}\ \bibnamefont {Sake}},\ and\ \bibinfo {author} {\bibfnamefont {S.~P.}\ \bibnamefont {Trivedi}},\ }\href {https://doi.org/10.1007/JHEP02(2024)145} {\bibfield  {journal} {\bibinfo  {journal} {JHEP}\ }\textbf {\bibinfo {volume} {02}},\ \bibinfo {pages} {145}},\ \Eprint {https://arxiv.org/abs/2307.15900} {arXiv:2307.15900 [hep-th]} \BibitemShut {NoStop}%
\bibitem [{\citenamefont {Cotler}\ and\ \citenamefont {Jensen}(2024)}]{Cotler:2024xzz}%
  \BibitemOpen
  \bibfield  {author} {\bibinfo {author} {\bibfnamefont {J.}~\bibnamefont {Cotler}}\ and\ \bibinfo {author} {\bibfnamefont {K.}~\bibnamefont {Jensen}},\ }\href {https://doi.org/10.1007/JHEP12(2024)016} {\bibfield  {journal} {\bibinfo  {journal} {JHEP}\ }\textbf {\bibinfo {volume} {12}},\ \bibinfo {pages} {016}},\ \Eprint {https://arxiv.org/abs/2401.01925} {arXiv:2401.01925 [hep-th]} \BibitemShut {NoStop}%
\bibitem [{\citenamefont {Anninos}\ \emph {et~al.}(2024)\citenamefont {Anninos}, \citenamefont {Baracco},\ and\ \citenamefont {M\"uhlmann}}]{Anninos:2024iwf}%
  \BibitemOpen
  \bibfield  {author} {\bibinfo {author} {\bibfnamefont {D.}~\bibnamefont {Anninos}}, \bibinfo {author} {\bibfnamefont {C.}~\bibnamefont {Baracco}},\ and\ \bibinfo {author} {\bibfnamefont {B.}~\bibnamefont {M\"uhlmann}},\ }\href {https://doi.org/10.1088/1475-7516/2024/10/031} {\bibfield  {journal} {\bibinfo  {journal} {JCAP}\ }\textbf {\bibinfo {volume} {10}},\ \bibinfo {pages} {031}},\ \Eprint {https://arxiv.org/abs/2406.15271} {arXiv:2406.15271 [hep-th]} \BibitemShut {NoStop}%
\bibitem [{\citenamefont {Held}\ and\ \citenamefont {Maxfield}(2024)}]{Held:2024rmg}%
  \BibitemOpen
  \bibfield  {author} {\bibinfo {author} {\bibfnamefont {J.}~\bibnamefont {Held}}\ and\ \bibinfo {author} {\bibfnamefont {H.}~\bibnamefont {Maxfield}},\ }\href@noop {} {\  (\bibinfo {year} {2024})},\ \Eprint {https://arxiv.org/abs/2410.14824} {arXiv:2410.14824 [hep-th]} \BibitemShut {NoStop}%
\bibitem [{\citenamefont {Buchmuller}\ \emph {et~al.}(2024)\citenamefont {Buchmuller}, \citenamefont {Hebecker},\ and\ \citenamefont {Westphal}}]{Buchmuller:2024ksd}%
  \BibitemOpen
  \bibfield  {author} {\bibinfo {author} {\bibfnamefont {W.}~\bibnamefont {Buchmuller}}, \bibinfo {author} {\bibfnamefont {A.}~\bibnamefont {Hebecker}},\ and\ \bibinfo {author} {\bibfnamefont {A.}~\bibnamefont {Westphal}},\ }\href@noop {} {\  (\bibinfo {year} {2024})},\ \Eprint {https://arxiv.org/abs/2412.09211} {arXiv:2412.09211 [hep-th]} \BibitemShut {NoStop}%
\bibitem [{\citenamefont {Honda}\ \emph {et~al.}(2024)\citenamefont {Honda}, \citenamefont {Matsui}, \citenamefont {Numajiri},\ and\ \citenamefont {Okabayashi}}]{Honda:2024hdr}%
  \BibitemOpen
  \bibfield  {author} {\bibinfo {author} {\bibfnamefont {M.}~\bibnamefont {Honda}}, \bibinfo {author} {\bibfnamefont {H.}~\bibnamefont {Matsui}}, \bibinfo {author} {\bibfnamefont {K.}~\bibnamefont {Numajiri}},\ and\ \bibinfo {author} {\bibfnamefont {K.}~\bibnamefont {Okabayashi}},\ }\href@noop {} {\  (\bibinfo {year} {2024})},\ \Eprint {https://arxiv.org/abs/2412.20398} {arXiv:2412.20398 [gr-qc]} \BibitemShut {NoStop}%
\bibitem [{\citenamefont {Iizuka}\ and\ \citenamefont {Sake}(2025)}]{Iizuka:2025vkl}%
  \BibitemOpen
  \bibfield  {author} {\bibinfo {author} {\bibfnamefont {N.}~\bibnamefont {Iizuka}}\ and\ \bibinfo {author} {\bibfnamefont {S.~K.}\ \bibnamefont {Sake}},\ }\href@noop {} {\  (\bibinfo {year} {2025})},\ \Eprint {https://arxiv.org/abs/2501.02614} {arXiv:2501.02614 [hep-th]} \BibitemShut {NoStop}%
\bibitem [{\citenamefont {Dey}\ \emph {et~al.}(2025)\citenamefont {Dey}, \citenamefont {Nanda}, \citenamefont {Roy}, \citenamefont {Sake},\ and\ \citenamefont {Trivedi}}]{Dey:2025osp}%
  \BibitemOpen
  \bibfield  {author} {\bibinfo {author} {\bibfnamefont {I.}~\bibnamefont {Dey}}, \bibinfo {author} {\bibfnamefont {K.~K.}\ \bibnamefont {Nanda}}, \bibinfo {author} {\bibfnamefont {A.}~\bibnamefont {Roy}}, \bibinfo {author} {\bibfnamefont {S.~K.}\ \bibnamefont {Sake}},\ and\ \bibinfo {author} {\bibfnamefont {S.~P.}\ \bibnamefont {Trivedi}},\ }\href@noop {} {\  (\bibinfo {year} {2025})},\ \Eprint {https://arxiv.org/abs/2501.03148} {arXiv:2501.03148 [hep-th]} \BibitemShut {NoStop}%
\end{thebibliography}%

\end{document}